\documentclass{aa}

\usepackage{txfonts}
\usepackage{graphicx}   
\usepackage{amsmath}    
\usepackage[utf8]{inputenc} 
\usepackage{array}
\usepackage{multicol,rotating} 
\usepackage{longtable} 
\usepackage[section]{placeins}
\usepackage{lscape}
\usepackage{subcaption}
\usepackage{multirow} 

\begin{document}

\title{Jupiter Trojans spectrophotometry using Gaia DR3 catalog} 

  \author{S. Fornasier
          \inst{1,2}
          \and
          N. El-Bez-Sebastien\inst{1}
           }

   \institute{LIRA, Université Paris Cit\'e, Observatoire de Paris, Universit\'e PSL, Sorbonne Universit\'e, CNRS, F-92190 MEUDON, France
              \email{sonia.fornasier@obspm.fr}
         \and
           Institut Universitaire de France (IUF), 1 rue Descartes, 75231 Paris CEDEX 05
             }

   \date{Received on June 2025, accepted for publication on 26 Aug. 2025}

\abstract{Jupiter Trojans are considered to be among the most pristine Solar System bodies. They may have originated from the Kuiper belt, and were captured by Jupiter during planetary migrations. }
{We present the spectral characterization of Jupiter Trojans using data from the third data release by the Gaia mission (DR3) spectral catalog and from the literature, the analysis of the spectral slope correlations with physical parameters, and the comparison with the outer Solar System bodies main properties.}
{The Gaia DR3 minor bodies spectral catalog comprises 478 Jupiter Trojans. Gaia spectrophotometry is available in 16 spectrophotometric points covering the 0.33-1.08 $\mu$m  range. In this sample, we conserve the Trojans having a signal-to-noise ratio (SNR) $>$ 20, and we visually inspect the quality of those with lower SNR. This process reduces the Trojans Gaia database to 320 objects. To enhance the statistical analysis, we have also included the visible spectra (in the $\sim$ 0.45-0.95 $\mu$m range) of Trojans available in the literature. The final dataset includes 519 Trojans, 291 in the leading swarm (L4) and 228 in the trailing swarm (L5), which we classified using the Bus-Demeo and Mahlke classification schemes.}
{The Trojan population is dominated by featureless asteroids with red spectral slopes belonging to the D-type, and an important fraction ($\sim$ 40\%) belong to the Z-class (in Mahlke taxonomy), characterized by very red slopes. The L4 swarm shows a higher spectral variability and a higher amount, by a factor of 2, of less spectrally red asteroids belonging to the C, P, and X classes, mainly associated with families members. 
Once excluding peculiar objects such as the members of Eurybates family, known to be dominated by C-type asteroids, the two swarms have indistinguishable average slope and very similar albedo values (7.86 $\pm$ 0.15\% and 7.35 $\pm$0.15\% for the L4 and L5 swarms, respectively). We do not observe a spectral  color bi-modality distribution,  conversely to the results previously reported in the literature from which it was suggested that Trojans originated from two different regions of the protoplanetary disk: one before and one behind the stability lines of some volatile ices, such as H$_{2}$S.The spectral slope distribution is peaked at 9-11 \%/1000 \AA, and it is very narrow compared to that of Transneptunian objects (TNOs) or cometary nuclei.}
{In the visible range, L4 and L5 Jupiter Trojans show very similar average slope and albedo values, even if the L4 swarm have a higher spectral variability mostly associated with some peculiar families. This points toward a common origin of the bodies of the two swarms, likely from the Transneptunian region.  Trojans visible spectral slopes match those of the less red TNOs, notably those classified as bowl-type by \cite{PinillaAlonso2024}, and lack of extremely red and organic rich surfaces observed in the TNOs.  We suggest that the Jupiter Trojans might have been captured by Jupiter from the Centaurs and scattered disk population, and that the lack of extremely red objects among Trojans is due to removal of the organic-rich crust through the sublimation of volatiles and collisions as TNOs migrated inward in the Solar System. }


   \maketitle

\section{Introduction}

Jupiter Trojans are  small bodies of the Solar System located in the leading (L4) and trailing (L5) Lagrange points of the Jupiter-Sun system, at a distance of 5.2 au from the Sun. The number of Trojans discovered to date is 15267 \footnote{\url{https://minorplanetcenter.org/iau/lists/JupiterTrojans.html}} of which 9684  belong to the L4 cloud and 5583 to the L5 one. The number of L4 Trojans with a radius greater than 1 km is estimated to be around 1.6 $\times$10$^{5}$ \citep{Jewitt2000}, comparable with the  main belt population of similar size. Trojans have orbits stable over the age of the Solar System \citep{Levison1997}, so their origin must date back to the early phase of the Solar System formation. \cite{Marzari_1998} hypothesized that they formed in close proximity to their present location and were subsequently trapped during the growth of Jupiter. \cite{Morbidelli_2005} proposed that Trojans originated in the Kuiper belt and were captured in the Jupiter L4 and L5 Lagrange points during the process of planetary migration, occurring shortly after the crossing of Jupiter and Saturn through their mutual 1:2 resonances. In either scenario, these primitive bodies are considered to be crucial in deciphering the history and the evolution of the Solar System.

The Trojan swarms contain dark and compositionally primordial asteroids. They are dominated by the spectrally red  D and P classes, following the Tholen classification scheme \citep{Tholen_1989}. Jupiter Trojan clouds are at least as collisionally evolved as main belt asteroids, and several dynamical families where identifies. \cite{fornasier_2004, fornasier_2007} characterized the composition of members of several dynamical families, discovering that the L4 Eurybates family is peculiar. In fact, it is dominated by carbonaceous C-type asteroids, while the majority of Trojans belongs to the D-type. These authors reported that, even if the two Trojans clouds are dominated by D-type asteroids, the L4 swarm shows an higher spectral variability with the presence also of objects with neutral to low spectral slope belonging to the carbon-rich C-type complex, findings confirmed by \cite{Roig_2008}. \cite{Emery_2011} identified a color bimodality  based on the near infrared colors (in the 0.7-2.5 $\mu$m range): a less red group, potentially originated from near Jupiter or in the asteroid main belt, and a redder group, probably originated from the outer Solar System. These authors did not find any albedo variation between the two groups.   

As Trojans form  beyond the frost line, they may be composed of organic compounds and may still contain ices in their interior, even if absorption features associated with water ice have not yet been clearly and univocally detected on their surfaces \citep{Emery_2004, Dotto_2006, Wong_2024, Brown_2025}. Their emissivity spectra indicated rather the presence of fine-grained silicates \citep{Emery_2006}. \cite{Brown_2016} discovered a 3.1 $\mu$m band on some Trojans, and this feature has been attributed to either fine-grained frost covering dark silicate grains or to N-H band produced by irradiation of ices containing some nitrogen. This last explanation is consistent with the notion that these objects originated beyond the giant planet region, where NH$_3$ ice was stable and produced non volatiles residues when space weathered, before reaching the Trojans swarms.

To investigate these primordial bodies, NASA launched in 2021 the Lucy mission. This mission is designed to make close approaches to four L4 Trojans (Eurybates, Polymele, Orus, and Leucus) of different spectral classes in 2027-2028, and one binary system (Patroclus-Menoetius) in the L5 swarm in 2033 \citep{Levison_2021}.\\
Recent observations with the JWST, focused on Lucy mission targets \citep{Wong_2024} and relatively bright Trojans \citep{Brown_2025}, reveal a complex and varied surface composition: Lucy targets exhibit a broad and deep (4-8\% depth) OH band centered at $\sim$ 3 $\mu$m, and an absorption in the 3.3-3.6 $\mu$m region attributed to aliphatic organics \citep{Wong_2024}. The OH band is deeper (4-8\%) in the spectrally less red targets Eurybates, Polymele and Patroclus, while the organic band is more prominent in the spectrally redder Orus and Leucus asteroids. Eurybates is unique not only for its moderate spectral slope, but it is the only Trojans showing a 4.25 $\mu$m band attributed to bound or trapped CO$_2$ \citep{Wong_2024}.  \\ The high albedo Trojans (p$_v \sim$ 10-13\%) investigated by \cite{Brown_2025} exhibit a moderate slope value and a broad  hydration feature centered at 3 $\mu$m,  with no discernible features attributable to ice. \cite{Brown_2025}  suggests that these objects may have originated from recent collisions and represent a third class of Jupiter Trojans alongside the red and less red classes identified by \cite{Emery_2011}.

In this paper we present the results of the spectrophotometric survey of Jupiter Trojans carried out in the 0.4-1 $\mu$m range with the Gaia mission. We combined the Gaia results with those available in the literature for a statistical analysis of the spectral behavior of the Trojans population in the visible range.  

\section{Data analysis}

\subsection{Gaia data}

Gaia is a cornerstone mission of the European Space Agency, launched in 2013.  The third data release (DR3) contains astrometric information as well as a minor bodies spectral catalog. This catalog includes the spectroscopy of 60,518 solar system objects \citep{tanga2023} measured with the Blue and Red Photometers (BR and BP), which cover the 0.33--0.68 $\mu$m and the 0.64-1.05 $\mu$m wavelength range, respectively.  The data reported in the DR3 were acquired between August 2014 and May 2017. In the catalog, each target's reflectivity is an average of multiple observations taken at different epochs during 2014-2017. \\
For each asteroid, the spectral observations are reduced to 16 spectrophotometric points, with each point representing the average reflectance over bins of 0.044 $\mu$m centered at fixed wavelengths (0.374, 0.418, 0.462, 0.506, 0.550, 0.594, 0.638, 0.682, 0.726, 0.770, 0.814, 0.858, 0.902, 0.946, 0.990, and 1.034 $\mu$m). The mean reflectance is weighted as: $$\bar{R}(\lambda)=\frac{1}{\sum w}\sum wR(\lambda)$$ where $w$ is the inverse of the uncertainty. The reflectance is normalized at 0.55 $\mu$m. For details on the Gaia DR3 asteroid spectrophotometry, we refer to the works of  \cite{tanga2023} and  \cite{galluccio2023}. \\

\begin{figure}[h!]
    \centering
    \includegraphics[width=0.99\linewidth]{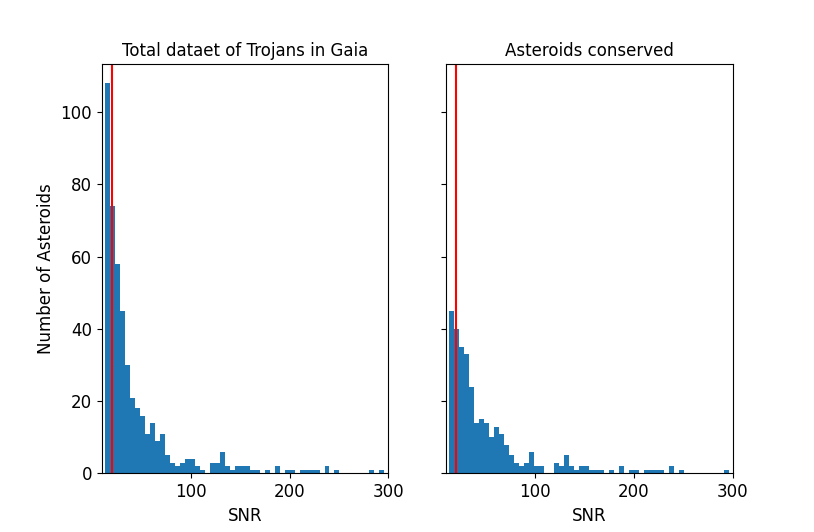}
    \caption{Histogram of the SNR distribution: on the left the one of all the Trojans present in Gaia's spectral catalog; on the right panel the one of the DR3 Trojans retained for this study. The red line corresponds to SNR=20. }
    \label{snrtrojans}
\end{figure}

In the DR3 catalog, 478 Jupiter Trojans were observed. We first inspected the quality of the data estimating their signal to noise ratio (SNR), following the definition reported in \cite{Tinaut-Ruano_2024}: 
\[ $$SNR =\frac{1}{N}\sum_{n=1}^N{R(\lambda_n)/R_{error}(\lambda_n})$$ \]
where $R(\lambda)$ is the normalized reflectance for a given wavelength $\lambda$ and $R_{error}$ the associated error reported for each spectrophotometric value \citep{galluccio2023}, and $N$ is the number of spectrophotometric points used in the SNR computation. We then systematically conserved all the Trojans with an SNR $>$ 20, which is considered a reliable threshold for asteroid spectral classification in the visible range \citep{Galinier_2024}. We visually inspected the spectral quality of those with a lower SNR. This inspection led to the removal of approximately 160 Trojans, for which the SNR is too low and the spectral behavior is unreliable (high peak-to-peak reflectance scatter from contiguous spectrophotometric points, or inconsistent spectral behavior between the BP and BR photometers). This reduced the dataset to 320 objects, 264 having SNR $>$ 20 and 56 with lower SNR (Fig.~\ref{snrtrojans}).

The spectrophotometric data are flagged 0, 1, and 2, indicating reliable, poor quality and untrustable data, respectively. We systematically exclude flag 2 points from the analysis and conduct a visual inspection of the quality of each flag 1 point. We remove them when their uncertainties are very high or when they are inconsistent with the overall spectral behavior. Additionally, the edges and overlapping regions between the two spectrophotometers are often affected by significant systematic errors, making them unreliable due to the low efficiency of the photometers \citep{Galinier_2024, galluccio2023}. For this reason, we systematically removed the 0.37 and 1.03 $\mu$m data,  and we carefully inspected the reflectance value at the overlapping wavelength between the BP and BR photometer,  discarding it in the SNR and spectral slope computation and taxonomic classification when its spectrophotometry was anomalous compared to the contiguous bands. \\
Finally, the data covering the UV/blue range were also corrected using the correction factors reported by \cite{Tinaut-Ruano_2023}. These authors identified an artificial spectral reddening (i.e. higher slope) in the DR3 data attributed to the solar analog stars used to generate the relative reflectance, stars which are not fully compliant with the solar spectrum in the 0.37-0.5 $\mu$m range. \\

We then proceed to the taxonomic classification using the Bus-Demeo and Mahlke classification scheme \citep{Demeo_2009, Mahlke_2022}, applying a chi-squared best fit between a given asteroid and the classes mean reflectance spectra from these taxonomies: \\
\[ $$\chi^2=\sum_{\lambda=0.42}^{\lambda=0.99}{\frac{{(R_{spectra}}_\lambda-{R_{class}}_\lambda)^2}{{\sigma_{class}}_\lambda^2}}$$ \]
where $R_{spectra}$ is the reflectance of a given asteroid spectrum,  $R_{class}$ and $\sigma_{class}$ are the mean reflectance and the associated uncertainty  for a given taxonomic class, respectively. \\
Finally, a visual inspection on the best fits results is performed to determine and to validate the final taxonomic classification of a given asteroid. \\
The Gaia spectrophotometry of the Trojans retained in the analysis is reported in Figs.~\ref{gaia_1},~\ref{gaia_2},~\ref{gaia_3}, and ~\ref{gaia_4}.

We computed the spectral slope between 0.550 $\mu$m and 0.814 $\mu$m. This range was selected because it is commonly adopted in the literature for spectral analysis of asteroids \citep{fornasier_2007} in the visible range, permitting the comparison of the Gaia spectrophotometry with the spectral slope values reported in the literature for Trojans and other outer solar System minor bodies.

\subsection{Literature data}

To enhance the statistical analysis, we have also considered spectra from the literature:  69 Trojans from \citet{fornasier_2007}, 16 from \citet{jewitt}; 117 are from photometric data from \citet{Roig_2008}; 17 from \citet{melita2008physical}; 13 from \citet{Laz}; 4 from \citet{Fitz}; 23 from \citet{Bendjoya}; and 35 from the Small Main-belt Asteroid Spectroscopic Survey (SMASS) I \citep{SMASS1} and SMASS II \citep{SMASS2}. The complete dataset of Trojans for which spectral or spectrophotometric data exists in the visible range comprises 519 objects, 291 in L4 and 228 in L5. \\
For Trojans with multiple observations, we averaged the spectral slope or eventually prioritized data with the higher SNR. Gaia DR3 provides new spectrophotometric data of 268 Trojans, that have not been previously observed in the literature. 

For the literature data, we took the spectral slope value, when published, or we computed it directly from the available spectra. In order to ensure consistent procedures and a homogeneous data set, the data were re-normalized at 550 nm if originally normalized at a different wavelength, such as those from \cite{Roig_2008}. 

The slope values for the entire Trojans sample here analyzed are reported in Tables~\ref{tab:t1} and ~\ref{tab:t2} (available at the CDS).

\section{Trojans spectral classification}

\begin{table*}
 \caption{Trojan's average spectral slope (in $ \%/1000 ~\AA $) for different intervals in size.}
    \label{tab:avg_slope_trojans}
    \centering
    \begin{tabular}{|c|c|c|c|c|c|}
    \hline
         Diam. & 0-25km & 25-50km & 50-75km &75-100km& $>$100km  \\
         \hline
        $\bar{S}$  & $8.63 \pm 0.27$ & $ 9.23 \pm 0.22$ & $9.22 \pm 0.37 $ & $8.42 \pm 0.73 $ & $7.27 \pm 0.71 $\\
        \hline
         $\bar{S}_{L4}$ &$7.35\pm 0.39$&$9.26 \pm 0.33 $&$9.22\pm 0.45$&$8.82\pm 0.94$&$8.04\pm 0.83$\\
         \hline
         $\bar{S}_{L5}$ &$10.03\pm 0.32$& $ 9.20 \pm 0.29$ & $9.21\pm 0.65$&$7.94\pm 1.19 $&$6.32\pm 1.17$\\
         \hline
    \end{tabular}
\end{table*}

\begin{table*}
 \caption{Spearman's rank correlation coefficients ($\rho$  and $P_r$) between slope and diameter, albedo and orbital elements.}
    \label{spearman}
 \centering
    \begin{tabular}{|c|c|c|c|}  \hline
      & Slope (all Trojans)         & Slope (L4)       &  Slope  (L5) \\  
      & $\rho$ ($P_r$)  & $\rho$ ($P_r$)   &  $\rho$ ($P_r$)   \\ \hline
Diam. (km)    &  0.0112 (0.7984) & 0.1410 (0.0156) & -0.1549 (0.0192) \\
p$_v$ & -0.1467 (0.0008) & -0.2285 (0.00008) & 0.0219 (0.7418) \\
e     & -0.0166 (0.7048) &  0.0272  (0.6428) & -0.0863 (0.1940) \\
a     & -0.0799 (0.0687) & -0.0623  (0.2893)  & -0.0908 (0.1717) \\
i     &  0.265684 (7.84$\times$10$^{-10}$)  & 0.3939 (3.04$\times$10$^{-12}$) & 0.0814 (0.2206) \\ \hline
 &  Diam. (km)  & Diam. (km)  & Diam. (km)   \\
  p$_v$    & -0.2587 (2.22$\times$10$^{-9}$) & -0.1889 (0.0012) & -0.3509 (5.24$\times$10$^{-8}$) 
\\ \hline
 \hline
    \end{tabular}       
\end{table*}

\begin{figure}
    \includegraphics[width=0.45\textwidth]{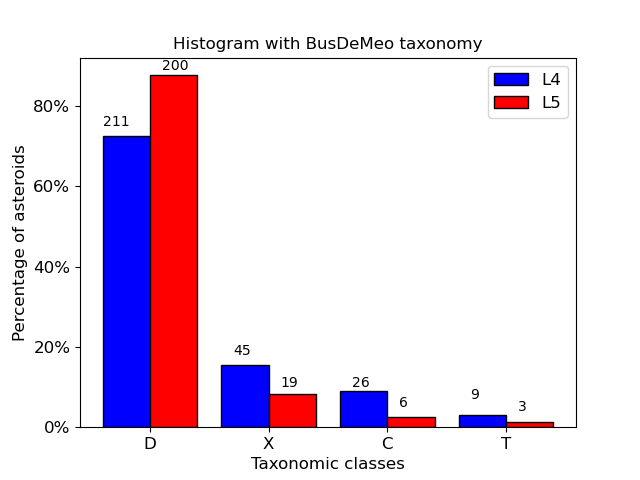}
    \includegraphics[width=0.45\textwidth]{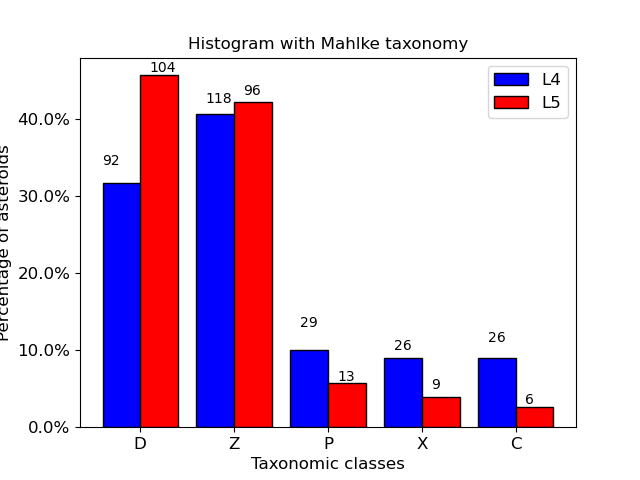}
    \caption{The histograms show the Trojan taxonomic classes identified using the Bus-DeMeo (upper panel) and Mahlke (lower panel) taxonomies \citep{Demeo_2009, Mahlke_2022}. The blue and red bars indicate the L4 and L5 asteroids, respectively.}
\label{histo_taxa}
\end{figure}
  
\begin{figure}
    \centering
    \includegraphics[width=0.8\linewidth]{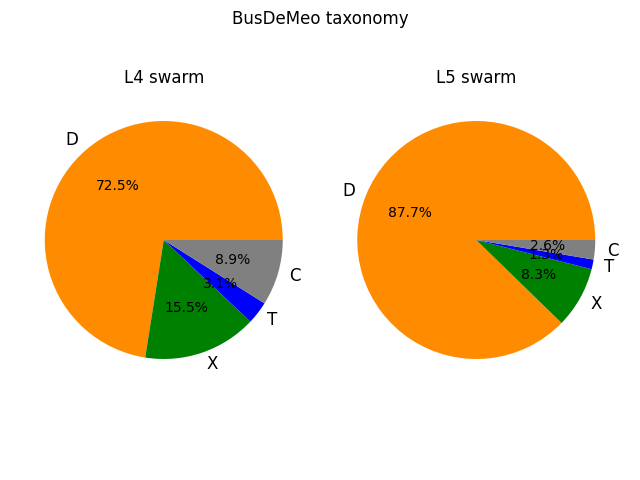}
    \includegraphics[width=0.8\linewidth]{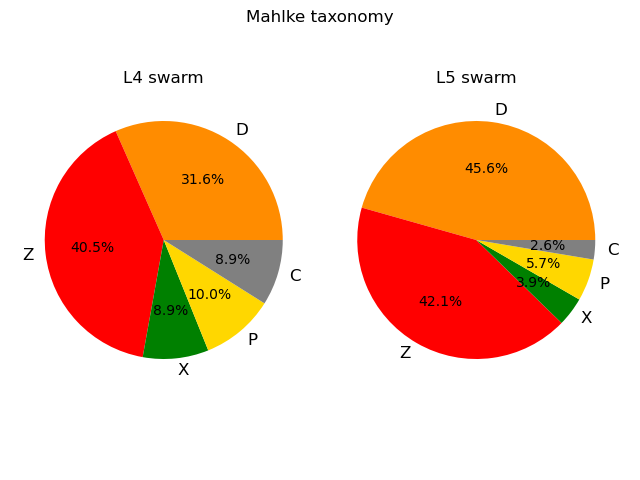}
    \caption{The pie charts show the distribution of the Trojan taxonomic classes according to the Bus-DeMeo (upper panel) and Mahlke (lower panel) classification schemes \citep{Demeo_2009, Mahlke_2022}. }
    \label{pie_taxa}
\end{figure}

\begin{figure}
    \centering
    \includegraphics[width=0.99\linewidth]{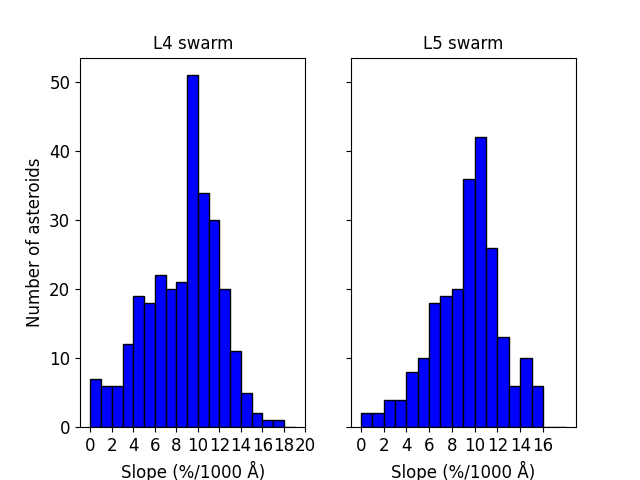}
    \includegraphics[width=0.99\linewidth]{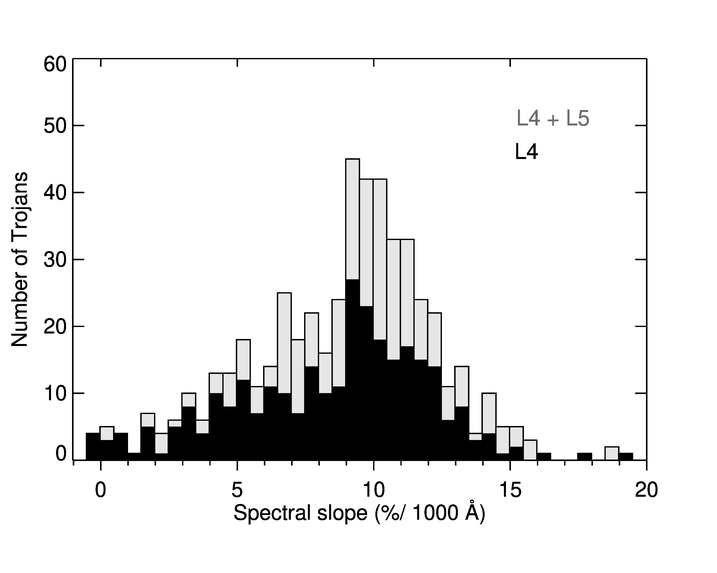}
    \caption{Top: histogram of the spectral slopes for the L4 and L5 swarms. Bottom: Histogram of the spectral slopes for all the Trojans investigated here (L4 are shown in black).}
    \label{hist_slope}
\end{figure}

The Trojan population is dominated by featureless (in the visible range) asteroids with red spectral slopes, as already reported in the literature \citep{fornasier_2007, Roig_2008}. In the Trojans sample here analyzed, according to the Bus-Demeo taxonomy \citep{Demeo_2009},  D-type asteroids dominate in both swarms (Figs~\ref{histo_taxa}, and \ref{pie_taxa}): 72.5\% in L4 and 87.7\% in L5. In addition to D-type, the L4 swarm also contains 15.5\% X-type, 3.1\% T-type, and 8.9\% C-type, while the L5 swarm has 8.3\% X-type, 1.3\% T-type, and 2.6\% C-type. Most of the X-type are featureless and dark asteroids that would be classified as P-type in the Tholen taxonomy \citep{Tholen_1989}. \\
The Mahlke taxonomy confirms the previous findings and allows to highlight the presence of very red asteroids, for which a new class, the Z-class, has been explicitly introduced (Figs~\ref{histo_taxa}, and \ref{pie_taxa}). In this taxonomy, the L4 swarm has 40.6\% Z-type, 31.6\%  D-type, 10.0\% P-type, and an equal percentage, 8.9\%, of X- and C-type asteroids. The L5 swarm has 45.6\% D-type, 42.1\% of Z-type, 5.7\%  P-type, 3.9\%  X-type and only 2.6\% of C-type asteroids. \\
The percentage of less spectrally red asteroids belonging to the C, P, and X classes is about a factor of 2 higher in the L4 swarm. This swarm shows a higher spectral variability, as noted by \cite{fornasier_2007}, including both spectrally flat C-type asteroids, mainly found in the very robust Eurybates family \citep{fornasier_2007, Roig_2008}, and very red Z-type asteroids, which dominate this swarm in percentage. \\

The spectral slope distribution is shown in Fig.~\ref{hist_slope}. On average, L4 asteroids have a spectral slope of $8.45\pm0.21 (\%/1000$ \AA), slightly less than that of L5 Trojans (9.41$\pm$0.21 \%/1000 \AA)), and their slope peaks between 9 and 11 \%/1000 \AA. The L4 asteroids here analyzed includes 88 families members as defined by \cite{BeaugeRoig2001}, and 202 background asteroids, while for the L5 swarm we have 68 families members \citep{BeaugeRoig2001} and 160 background asteroids. Excluding the asteroids members of families, the average spectral slope of L4 and L5 swarms are the same: 9.37$\pm$0.22 \%/1000 \AA\ for L4, and 9.34$\pm$0.24 \%/1000 \AA\ for L5.  \\
In the literature, spectroscopic studies of Trojans have been focused mainly on family members \citep{fornasier_2004, fornasier_2007, Dotto_2006, Roig_2008, deLouise_2010}. \cite{fornasier_2007} examined 142 objects, 74 from the L4 swarm and 68 from the L5 swarm, mostly members of different families, and found an average spectral slope of 4.57$\pm$4.01 \%/1000 \AA\ and 8.84$\pm$3.03 \%/1000 \AA\ for the L4 and L5 swarms, respectively. It should be noted that the low averaged spectral slope of the L4 swarm is strongly biased by the Eurybates family, which is the only family dominated by flat asteroids, or having moderate slopes values, such as C- and P-type (in Mahlke taxonomy). Excluding Eurybates members, these authors reported an average slope of 7.33 $\pm$ 4.24 \%/1000 \AA\, which is closer to the one we found. However, even excluding Eurybates, \cite{Roig_2008} and \cite{fornasier_2007} reported that L5 swarm has a higher average slope than L4, and contains a larger proportion of D-type than L4. Our analysis on a larger sample confirms that L5 Trojans have a slightly higher slope value. However, if we consider only background objects, there is no difference in the spectral slope between asteroids of the two swarms. 
This indicates that the L4 and L5 Trojans have a common origin and a similar evolution under irradiation, while families members, generated by collisions in the past hundreds million years, display a higher spectral variety including bodies having small spectral slope values, as the Eurybates family members. \\

Previous studies reported a bimodal distribution of spectral slope \citep{Szabo_2007, Roig_2008, Wong_2015, Tinaut-Ruano_2024, Emery_2011, Wong_2014, Emery_2024}, both in the visible (on a sample of almost 300 Trojans from the Sloan Digital Sky Survey (SDSS) colors \citep{Szabo_2007}, and NIR range (sample of 68 Trojans).  Two groups were distinguished, a red and a less red one.  Objects of the reddest group in the visible range appear consistently redder also in the near infrared range, up to 3.4 $\mu$m  \citep{Emery_2024}.  However, this bimodality is not observed in the largest sample of Trojans here investigated (Fig.~\ref{hist_slope}): there is no dip in the frequency of Trojans with slopes in the $\sim$ 6-8 \%/1000 \AA\ range, as reported from SDSS observations (\cite{Wong_2014}, their Fig. 1), but a continuously increasing slope value distribution with a peak at 9-11  \%/1000 \AA. Based on the previously reported bimodality  from SDSS observation,  it has been suggested that the two Trojans groups came from two different sources in the Solar System \citep{Wong_2016} located one before and one behind the stability lines of some volatile ices, such as H$_{2}$S, or that they have different regolith properties, and/or distinct surface composition \citep{Emery_2024}. Considering that we do not find any bimodal distribution in colors, that the two swarms have very similar albedo and slopes values, once family members excluded, and that the smaller Trojans show the higher color variability, we do not support the hypothesis that the Trojans are originated from different sources in the Solar System. We propose instead that the observed color distribution likely results from evolutionary processes like space weathering, erosion and removal of the organic rich refractory crust by collisions and/or eventually by cometary like activity, even if unlikely at those distances. 

This study, including 268 new spectrophotometric data on Trojans never observed before,  highlights a significative abundance of asteroids belonging to the very red Z class. The abundance of Z-type we found is much higher than that reported by \cite{Mahlke_2022} on their study, based on a sample of 86 Jupiter Trojans. These authors found that only 8\% of the Trojans in their sample belong to the Z-type, while the dominant class is D-type (51\%), followed by P-type (20\%), and M-type (12\%). Considering the main belt asteroids, \cite{Mahlke_2022} found that the Z-type represented only 1.1\% of a sample of 2125 bodies, while the D-type represented the 3.9 \%. Very few Z-type asteroids with a very steep spectral slope were observed in the main belt: (203) Pompeja and (269) Justitia, proposed by \cite{Hasegawa_2021} to be captured Transneptunians, or (72) Feronia and (732) Tjilaki, also proposed to have been originated at further heliocentric distances than the present ones \citep{Jules_2022}.

\section{Correlations between the spectral slope and physical parameters}

We look for correlations between the spectral slope and physical parameters of the Trojans. We ran a Spearman rank correlation between different parameters, which returns the rank correlation coefficient ($\rho$) and the significance of its deviation from zero ($P_r$). If $\vert \rho \rvert$ is close to zero this implies no correlation, if $\vert \rho \rvert$ is close to 1 the data are fully correlated, and intermediate values are usually considered as strong (0.7-1), moderate (0.5-0.7), and weak (0.2-0.5) correlation, respectively. The same limits, in absolute value, apply to anticorrelations when the coefficients are negative. $P_r$ varies between 0 and 1, and a correlation is significant when this coefficient is close to zero.

\begin{figure}
    \centering
    \includegraphics[width=0.99\linewidth]{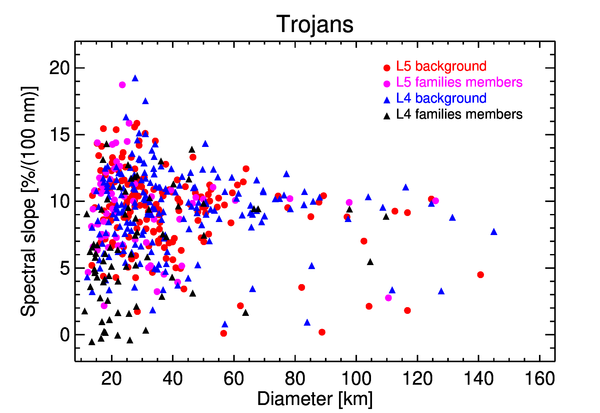}
    \caption{Spectral slope versus diameter for the Jupiter Trojans.} 
    \label{diamslope}                  
\end{figure}

The spectral slope versus diameter is shown in Fig.~\ref{diamslope} and in Table~\ref{tab:avg_slope_trojans}, for different size ranges. Smaller asteroids have a broader spectral slope distribution and a greater spectral variety than the larger bodies. The asteroids of the two swarms show no difference in the spectral slope in the 25-100 size range. For the Trojans smaller than $D< 25$ km, the L4 asteroids have a much lower spectral slope than the L5. However, this is due to the family members, which have an average slope of 5.23$\pm$0.25 $\%/(1000 ~\AA)$, determined in particular by the Eurybates family (most of the black triangles in the -2 -- 5 $\%/(1000 ~\AA)$ spectral range in Fig.~\ref{diamslope} are Eurybates members). If only the small L4 background asteroids are included, then the average slope is  9.01$\pm$0.40  $\%/(1000 ~\AA)$, much closer to the L5 bodies of similar size. In any case, the L5 Trojans smaller than 25 km are on average slightly redder compared to the other larger Trojans. For diameter larger than 100 km, the statistic is limited because there are only 9 L5 and 11 L4 members.  They clearly lack of extremely red asteroids, encountered mostly for Trojans smaller than 50 km. The less red colors might be due to differences in surface texture (grain size, roughness), or to diverse composition and  irradiation effects. \\
It should be noted that the completeness of the Trojan sample investigated here is likely not uniform. The selection criterion based on the quality of the spectra and the signal-to-noise ratio excludes the fainter objects in Gaia DR3. Moreover, this mission observed only $\sim$ 3.1\% of the known Trojans, likely missing the fainter ones. The smallest asteroids investigated here have estimated diameters of 11.2 km and 12.3 km for the L4 and L5 swarms, respectively (Figs~\ref{diamslope} and ~\ref{dia_alb_slope}). There are 92 Trojans with sizes smaller than 25 km in the L4 swarm and 84 in the L5 swarm. Since these smaller asteroids have similar heliocentric distances and comparable average albedo values (9.6$\pm$0.3\% for L4 and 8.4$\pm$0.3\% for L5), we do not expect significant selection bias when comparing the properties of the L4 versus L5 swarms.

\begin{figure}
    \centering
    \includegraphics[width=0.99\linewidth]{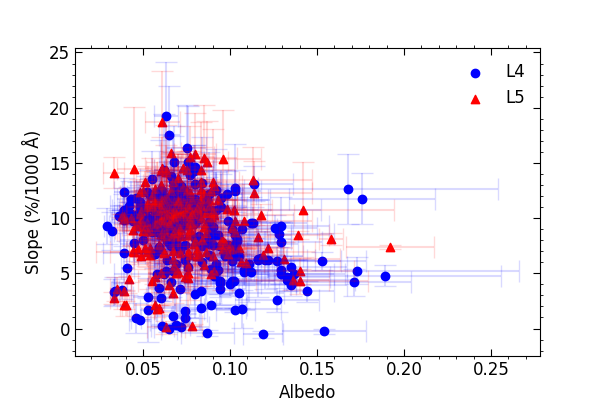}
    \caption{Spectral slope of the Trojans versus albedo. The L4 and L5 Trojans are represented by blue circles and red triangles, respectively.}
    \label{albslope}
\end{figure}

\begin{figure}
    \centering
    \includegraphics[width=0.99\linewidth]{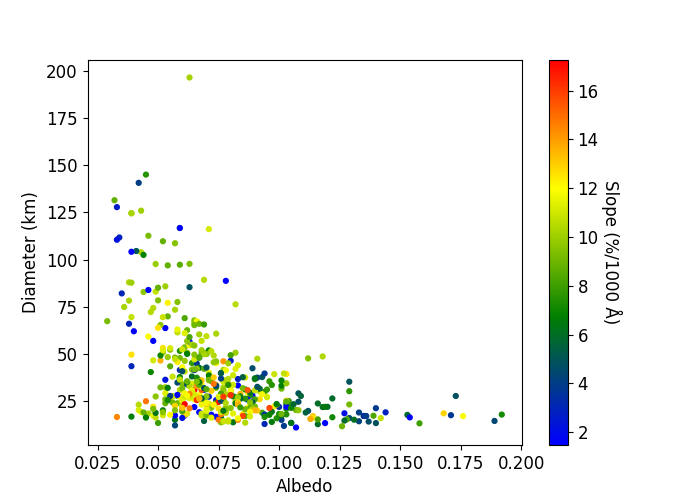}
    \caption{The diameter versus albedo of the Jupiter Trojans investigated here. The color of the data points is proportional to the spectral slope value.}
    \label{dia_alb_slope}
\end{figure}

The mean albedo is $7.86 \pm 0.15 \%$ for the L4 Trojans, and $7.35 \pm 0.15\%$ for the L5 ones. The whole Trojan sample here investigated  has an albedo of $7.62 \pm 0.12\%$, ranging from 2.9\% to 19.2\%. A subset of asteroids demonstrated anomalously high albedos for D-Z types, albeit with considerable uncertainty (Fig. ~\ref{albslope}). A comparative analysis of the mean albedos of the two swarms reveals no statistically significant differences. \\ 
We notice a very weak anti-correlation between the albedo and the spectral slope for L4 Trojans (Table~\ref{spearman}, and Fig.~\ref{albslope}), a finding that has not been previously documented in the literature \citep{emery2011, melita2008physical}. The anticorrelation remains consistent even when the members of the Eurybates family are excluded ($\rho$ =  -0.2), and it is particularly weak for background L4 Trojans ($\rho$ = -0.1127, $P_r$=0.1096). This finding indicates that L4 family members are slightly brighter and spectrally bluer than background Trojans. Rejuvenating processes following collisions may explain this behavior. 

A more significant anticorrelation is found between the albedo and the size (Table~\ref{spearman}), this time stronger for the L5 swarm (Fig.~\ref{dia_alb_slope}). Therefore, higher albedo and redder Trojans are predominantly observed for diameters smaller than 30-40 km, while the largest asteroids tend to exhibit darker and less red reflectance, and lack very red and bright objects. The fact that smaller Trojans show higher albedo values may be due to observational bias because the brightest asteroids are easily detected. However, among asteroids smaller than 25 km, we found bodies with both very low and relatively high albedo values. If larger, brighter asteroids exist, we expect them to be easily detected by observational surveys, including Gaia. In the present sample, all Trojans with a diameter greater than 50 km have an albedo value $<$ 8.2\%, and the mean albedo value of both the L4 and L5 members larger than 50 km is 5.5\%. Therefore, the absence of bright bodies among the larger Trojans is real and not associated with observational biases. Collisions among the smaller Trojans may easily expose brighter subsurface material, while space weathering processes or different surface textures may darken the larger Trojans.

The findings of our research have revealed no statistically significant correlations between the spectral slope and the semi-major axis or orbital eccentricity. A weak yet statistically significant correlation is observed between the slope and the inclination of the L4 asteroids (Table~\ref{spearman}), indicating that redder objects tend to have a larger inclination.   \citet{fornasier_2007}  observed a similar phenomenon but noted that this correlation was predominantly influenced by Eurybates family.  \citet{emery2011} found a weak color-inclination correlation for the less red  group of Trojans that they defined in their study, suggesting that that low-inclination Trojans may be bodies that orbited close or nearby Jupiter, and captured by the giant planet by gas drag. \citet{Roig_2008} found a slope-inclination correlation for the background asteroids of both  L4 and L5 swarms. In the dataset presented here, the correlation remains consistent even when the Eurybates members are excluded, although to a lesser extent, though still statistically significant ($\rho$=0.336 and P$_r$ = 1.176 $\times 10^{-8}$). Contrary to the results of \cite{Roig_2008}, the background Trojans examined here demonstrate the slope-inclination correlation exclusively in the L4 swarm ($\rho$=0.284 and P$_r$ =  4.052$\times 10^{-5}$), which is significantly more populated that the trailing one by a factor of 1.6 \citep{Szabo_2007}.

Our analysis indicates that redder L4 Trojans exhibit a higher inclination, while gray or moderately red Trojans are predominantly found at low inclination. This cannot be attributed exclusively to dynamical scattering resulting from collisions (therefore mostly observed in families members), as the observed correlation is also present in the background L4 objects. 

\section{Lucy's targets}

\begin{figure}
    \centering
    \includegraphics[width=0.99\linewidth]{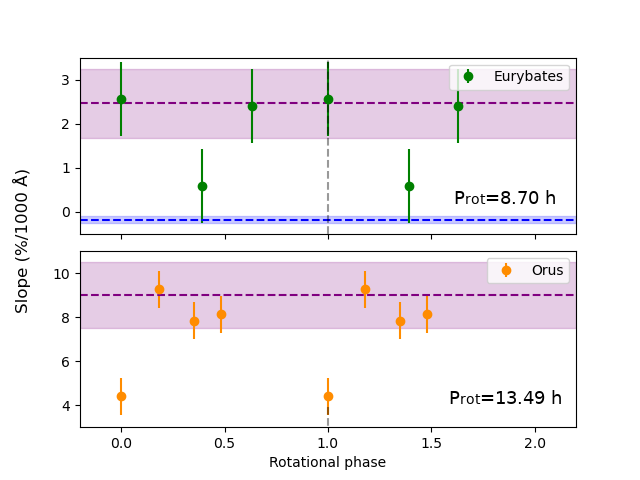}
    \caption{Spectral slope versus rotational phase for Eurybates and Orus. The purple area and the dashed line represent the spectral slope value with its uncertainty, as measured by Gaia. The spectral slope values derived by \citet{Souza_Feliciano_2020} are represented in green and orange for Eurybates and Orus, respectively. In the top panel the blue region represents the spectral slope obtained by \citet{fornasier_2007}.}
    \label{rotphase}
\end{figure}

In the Gaia DR3 catalog, spectrophotometric data are available for all the Lucy targets: (617) Patroclus-Menoetius, (3548) Eurybates, (11351) Leucus, (15094) Polymele, and (21900) Orus. Leucus and Polymele are both classified as D-type \citep{Demeo_2009}, but in both cases the Gaia spectra are very noisy with SNR $\le$20.
For Leucus, the spectral slope derived from Gaia is 12.04$\pm$3.40 \%/1000 \AA, in agreement within the large uncertainties with results from the literature (10.27$\pm$0.17 \%/1000 \AA, \cite{fornasier_2007}). Polymele belongs to the less red group of Trojans, with a slope from Gaia of 7.59$\pm$3.43 \%/1000 \AA. Patroclus-Menoetius has a high SNR and a spectral slope of 4.25$\pm$ 0.25. It also belongs to the less red group of Trojans, and it is classified as X-type in the Demeo taxonomy \citep{Demeo_2009}. \\
Eurybates was observed 59 times between 2014 and 2019, and the average DR3 spectrum of these observations is nearly flat with a moderate spectral slope of 2.46$\pm$0.79 \%/1000 \AA. We classified it as a C-type, confirming the taxonomy reported in the literature \citep{fornasier_2007, Souza_Feliciano_2020}. Eurybates shows some spectral variability: \citet{fornasier_2007} reported a slope of -0.18$\pm$ 0.57 \%/1000 \AA\  from observations acquired on 2003 at the VLT telescope, while  \citet{Souza_Feliciano_2020} observed it in 2018 at four different rotational phases (the Eurybates period is of  8.702724$\pm$0.000009 h \citep{Mottola_2016}), finding a spectral slope of about 2.5 \%/1000 \AA\ for 3 observations and 0.5 \%/1000 \AA\  for the remaining one (Fig. ~\ref{rotphase}).  Following the implementation of phase reddening corrections and the calculation of the slope within the same range, \cite{Souza_Feliciano_2020} observed consistent values between their lower slope value and the one reported by \cite{fornasier_2007}. Consequently, Eurybates exhibits spectral heterogeneities. The observed variation in the spectral slope of Eurybates is analogous to that observed for the eponymous family members (-0.5 to 4.6 \% /1000 \AA).\\  
The Gaia data are consistent with the highest slope value reported by \citet{Souza_Feliciano_2020}. The spectral variability of Eurybates, the largest member (D = 64 km) of the eponymous family, may be associated with impacts and cratering processes \citep{Souza_Feliciano_2020}, exposing locally fresher and bluer material. Furthermore,  Eurybates has a known satellite, Queta (D= $1.2 \pm 0.4$, \citet{eury_sat}). Its passage in front of the primary may originated some spectral variability. However, the flux contribution of the secondary object to the entire system is only of 0.04\%, assuming similar albedo values between the primary and the secondary. Therefore, we favor local heterogeneity in composition as source of the observed spectral variability.  The in-situ observations of Lucy will provide valuable insights into the surface composition and heterogeneity of this peculiar Trojan.  \\
The Gaia spectrophotometry of Orus yielded a spectra slope of  9.02$\pm$1.50 \%/1000 \AA, confirming its D-type classification. This target, which has a rotational period of 13.48617$\pm$0.00007 h \citep{Mottola_2016}, was also observed at different rotational phases by \citet{Souza_Feliciano_2020}. Three out of four of their observations give a high spectral slope value, consistent, within the uncertainties, with the one derived from Gaia observations. One observation has a spectral slope value lower by a factor of 2 compared to the other ones. However, \citet{Souza_Feliciano_2020} report problems in the instrumental setup and also seeing degradation during this acquisition, and they suggest to discard this data point. Excluding this point, the ground-based and Gaia observations exhibit comparable spectral slopes, suggesting a relatively uniform surface for this Lucy mission target.

\section{Comparison with other Solar System small bodies}

\begin{figure}
    \centering
    \includegraphics[width=1.0\linewidth]{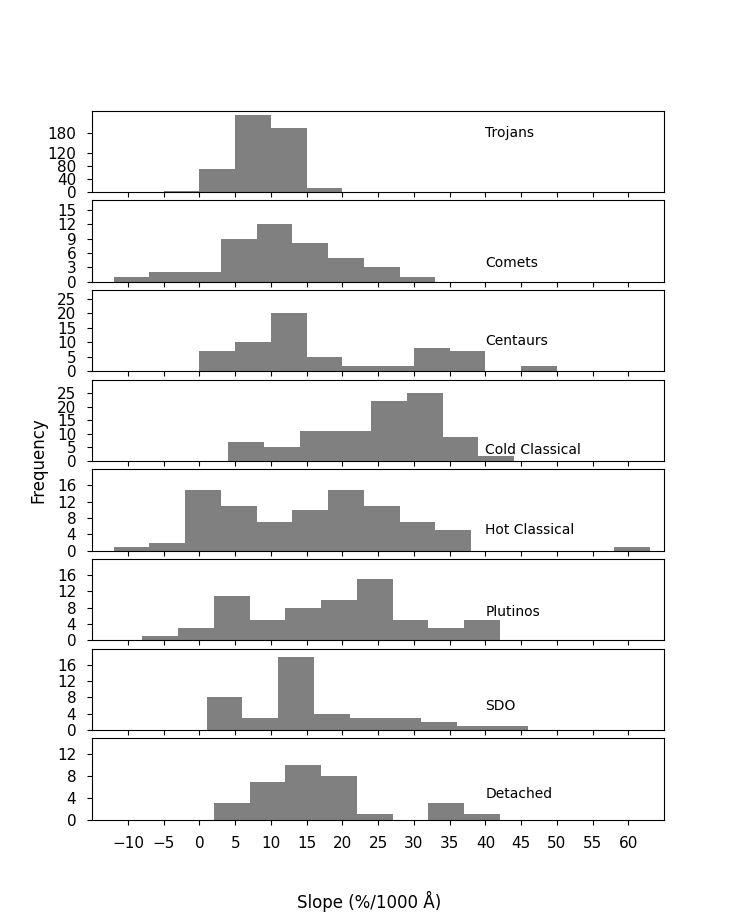}
    \caption{Comparison of the slopes of the Trojans (top plot) with different TNOs dynamical classes and cometary nuclei. Frequency indicates the number of asteroid bodies for each slope bins (the bin is of 5 \%/1000 \AA). From second top plot to bottom the histograms represent: Comets, Centaurs, Cold Classical, Hot Classical, Plutinos, Scattered Disk Objects (SDO), and detached TNOs.}
    \label{slopestno}
\end{figure}

Given the potential for Trojans to be Transneptunian Objects (TNOs) captured by Jupiter, we compared the spectral slope distribution of Trojans with that of various dynamical classes of TNOs and cometary nuclei in Fig.~\ref{slopestno}. These values were derived from those reported by \citet{Fornasier2009} and the Minor Bodies in the outer Solar System (MBOSS) catalog \citep{MBOSS} for a total TNOs sample of 170 objects. The slope distribution of the Trojans is similar to that of the less red population of TNOs and centaurs, i.e., bodies with slopes between 0 and 20\%/1000 \AA, which will be referred to as the "gray" population. Trojans are peculiar in that their slope distribution is much narrower than that of other outer solar system bodies, as previously noted by \cite{fornasier_2007}. In fact, the range of observed spectral slopes for TNOs and centaurs is much broader, including both bodies with flat spectra and the reddest ones in the solar system. The spectral distribution of the Trojans is similar to the "gray" population of Centaurs (with slopes of less than 18-20\%/1000 Å), the scattered disk objects (SDOs), and the detached population (i.e., TNOs having orbits whose perihelia are sufficiently distant from the gravitational influence of Neptune), but, as much narrower.

Looking at reflectance properties, the mean albedo of the L4 and L5 Trojans ($\sim$7.6\%) is very similar to that reported for the scattered disk objects (7.5$\pm$1.0 \%) and the Centaurs (7.4$\pm$0.6\%, \cite{Muller_tno}). It is also close to that of the hot classicals (8.4$\pm$1.4 \%), excluding the brightest dwarf planets and the Haumea family members (Fig.~\ref{albedoslopeall}). Trojans surfaces are much darker than those of cold classicals, outer resonant, or detached TNOs \citep{Muller_tno}. The average albedo of Trojans is comparable to, or slightly higher than that of cometary nuclei, which ranges between 3 and 7\% \citep{Filacchione_2024, Lamy2004}. \\
Considering both the visible spectral slope and the albedo, in the hypothesis that the Trojans are captured TNOs, the Centaurs and SDOs have the closest albedo values and spectral slope distributions, at least for their gray populations.  \\

\section{Discussion}

\begin{figure}[h!]
    \centering
    \includegraphics[width=1.0\linewidth]{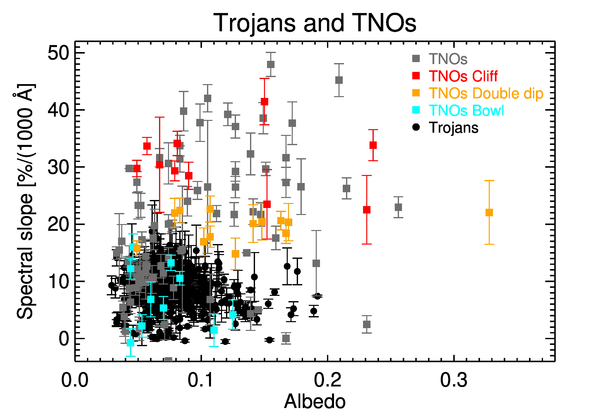}
    \caption{Spectral slope (evaluated in the visible range) versus albedo for the Trojans (black circles) and Transneptunians (gray square). The sub classes defined by \cite{PinillaAlonso2024} on the base of the 3 $\mu$m region features from JWST observations are represented in cyan (bowl type), orange (double dip type) and red (cliff type).}
    \label{albedoslopeall}
\end{figure}

Trojans are a peculiar population in the outer solar system: their spectral distribution is much narrower than that of other outer solar system bodies,  they lack the extremely red objects observed in the TNOs, and share similar albedo values to Centaurs, SDOs and cometary nuclei. To further explore connection with TNOs, in Fig.~\ref{albedoslopeall} we plot the visible spectral slope versus albedo for Trojans and TNOs. In this plot we distinguish the three types of TNOs defined by \cite{PinillaAlonso2024} based on the spectral characteristics of the 2.5-4 $\mu$m region from JWST data: bowl, double dip and cliff.  Although these types are related to features observed in the NIR region, we investigate their potential relationship with Trojans based on visible spectral slope and albedo. In addition to the previously noted similarity between Trojans and Centaurs and SDOs, Figure~\ref{albedoslopeall} shows that the spectral type of TNOs closest to Jupiter Trojans is the bowl type. The TNOs belonging to this class have spectra characterized by a deep, Gaussian-like shaped 3 $\mu$m band, together with the 1.5 and 2 $\mu$m bands, associated with water ice. The 4.27 $\mu$m band of CO$_2$ is also observed, while they lack organic and methanol features \citep{PinillaAlonso2024, Licandro2025}. This class is characterized by a low albedo and lower spectral slope in the visible range compared to the other two classes. \\
A few double-dip TNOs match the higher values  of the Trojans spectral slope. These TNOs have the highest abundance of CO$_2$ ice, but also contain water ice in smaller amounts than on the bowl type, and organics. Their visible spectral slope is  comprised between $\sim$ 17 and 22 \%/(1000 \AA), and they show a drastic flattening of the spectral slope starting from $\sim$ 1.2 $\mu$m. Cliff type are the reddest TNOs in the visible range, and have a number of features in the NIR range, mainly attributed to complex organics and methanol. \\
While the visible spectral behavior and albedo of Jupiter Trojans match those of bowl-type TNOs, their NIR behavior is completely different. The Trojans that have been investigated by spectroscopy, including the recent JWST data, have no clearly detectable water ice absorption features \citep{Dotto_2006, Emery_2004, YangJewitt_2007, Wong_2024, Brown_2025}, indicating that the surface water ice fraction should be negligible, conversely to TNOs bowl-type. Trojans spectra have very few absorption features and less prominent than TNOs : they show a broad OH band centered at 3 $\mu$m, an absorption in the 3.3-3.6 $\mu$m region due to aliphatic organics, and, in the case of Eurybates, a 4.25 $\mu$m band likely due to CO$_2$ clathrates \citep{Wong_2024}, considering that CO$_2$ ice is not stable on Trojans surfaces. 

The fact that Trojans lack very red surfaces as those observed in the cliff type TNOs sheds light on the composition and the surface evolution of outer Solar System bodies. Cliff type TNOs are richer in organics and methanol, which is known to be a high reddening agent \citep{Brown_2012, Brunetto2025}. Redder TNOs are thought to have formed beyond the methanol ice line during the early stages of solar system formation \citep{Brown_2012, Lacerda_2014}. 
The highest content of organics is in particular observed in TNOs having high eccentricity orbits \citep{Brunetto2025}, which exposes their surfaces to high irradiation doses. According to experimental simulations made by \cite{Quirico_2023}, irradiation of methanol ice produces organic materials containing aliphatic,  olefinic, acetylinic, carbonyl and hydroxyl groups, which are characterized by very red spectra.  They proposed that the combined action of space weathering, irradiation, and sputtering produces an organic-rich crust in the first millimeters of TNOs surfaces. This is followed by a subsurface layer of organics and ices at a depth of about 1 m, and a composition mostly made of ices at higher depths. The fact that Trojans do not display very red surfaces as those observed in the TNOs population might simply be due to the progressive erosion of the highly irradiated, organic- and methanol-rich crusts by sublimation of volatiles at low heliocentric distances. Centaurs, which are a transitional population between TNOs and comets, exhibit a color dichotomy with a redder and brighter group, rich in ices and organics, and a gray group that is darker and has surfaces richer in refractories and depleted in ices \citep{duffard_2014}.  The gray group is believed to be the result of higher cometary activity \citep{duffard_2014}. This is supported by the observation that they spent more time in orbits with smaller perihelion distances compared to the red group \citep{melita2012}. Interestingly, \citet{Belyakov_2024} find analogies between the spectral slope of Jupiter Trojans and that of low-perihelion TNOs.
 
Jupiter Trojans are expected to preserve water ice in their interiors. According to thermal models, water ice should be stable in the Trojans' subsurface (at a depth of about 10 meters, or less), but it should rapidly sublimate upon exposure to the surface \citep{Guilbert_2014}. Similarly to what observed on Trojans, ice features are rarely detected on the nuclei of comets \citep{Sunshine_2006, Barucci_2016, Filacchione_2016, Filacchione_2024} , which are known to be rich in volatiles. This is because the dark refractory coating optically dominates and masks volatile features.  \\
Moreover, the Rosetta mission's detailed investigation of comet 67P has proven that the typical size of ice exposures on cometary nuclei is about a meter or less \citep{fornasier_2023}, and that ice is often exposed in the form of frost with a very short lifetime \citep{deSanctis_2015, Fornasier_2016}. Therefore, high spatial and temporal resolution is needed to unambiguously detect ice on primordial dark bodies. Even if water ice is not directly exposed to the surface, there are several evidences that it is abundant in the subsurface and generates significant color changes in the nucleus due to the diurnal and seasonal cycles of water \citep{deSanctis_2015, Fornasier_2016, Filacchione_2024}. \\
Trojans are very likely a population in between TNOs, Centaurs and cometary nuclei. Their surface composition and properties should have been shaped by the sublimation of hypervolatile ice during their inward migration through the solar system, as well as by space weathering and collisions. \\
The Lucy mission offers a unique opportunity to study different Trojans in-situ, enabling an in-depth investigation of various bodies and shedding light on their nature and composition.

\section{Conclusions}

We have presented the results of the taxonomic classification and spectral slope analysis of the largest sample of Jupiter Trojans investigated to date, comprising 519 objects. The sample includes primarily spectrophotometric data from the Gaia DR3 spectral catalog, as well as data from the literature. Notably, it includes 268 new spectrophotometric data on Trojans that have never been observed before. Our main findings are as follows:\\
\begin{itemize}
\item  Red asteroids belonging to the D-type (in the Bus-Demeo taxonomy) dominate both Trojan swarms, accounting for 73\% and 88\% of the L4 and L5 populations, respectively. We found a significant proportion (40\%) of highly red Trojans belonging to the Z-type in the Mahlke taxonomy.
\item The L4 swarm shows higher spectral variability, with a proportion of C-, P- and X-type asteroids two times higher than that of the L5 Trojans. This variability is mostly associated with family members, notably the peculiar Eurybates family, and does not appear to be intrinsic, but rather the result of collisional processes that generate families. Additionally, smaller Trojans (with a diameter $<$ 40 km) exhibit greater spectral variability, whereas larger ones tend to lack very red surfaces and are darker.
\item Once the family members are excluded, L4 and L5 asteroids have very similar albedo (7.6 \%) and visible spectral slope values (9.3 $\%/ 1000 ~\AA$), indicating a common origin  and a similar evolution under irradiation processes over time.
\item We look for correlations between the spectral slope and orbital and physical parameters, finding : a weak anticorrelation between the spectral slope and the albedo for L4, which is stronger for family members and is likely attributable to rejuvenating processes due to collisions;    a correlation between the spectral slope and the inclination for the L4 Trojans ; an anticorrelation between the spectral slope and the albedo for the L5 Trojans.
\item Conversely to the results reported in the literature, we do not find a bimodal spectral slope distribution for the Jupiter Trojans but an asymmetric distribution that peaks at 9-11 $\%/ 1000 ~\AA$.   Our analysis does not support the conclusion, based on the previously reported bimodal slope distribution, that Trojans originate from two different sources in the Solar System. Instead, we favor a common origin and evolutionary processes, such as collisions, as the cause of the observed spectral variability.
\item  Even though Trojans exhibit spectral variability, their spectral distribution is narrower than that of other outer Solar System bodies, as has been noted in previous literature.
\item The gray groups of Centaurs and Scattered TNOs are closest to Jupiter Trojans in terms of spectral slope and albedo, and therefore potential sources of Jupiter Trojans
\item The visible spectral slope of Trojans is similar to that of water ice-rich TNOs belonging to the bowl type, as defined by \cite{PinillaAlonso2024}. However,  the NIR spectra of Trojans do not show evidence of water ice absorption features. If water ice is present, it must be in a very low quantity on the surface.  
\item Trojans lack the extremely red colors observed on TNOs, which are associated with organics rich surfaces  resulting from the irradiation of volatile ices, such as methanol ice. Assuming that Trojans originate from the Kuiper belt, we interpret the absence of extremely red bodies among Jupiter Trojans as resulting from the removal of the organic-rich layer through the sublimation of volatiles and collisions.

\end{itemize}

\section{Data availability}

Tables~\ref{tab:t1} and ~\ref{tab:t2} are only available in electronic form at the CDS via anonymous ftp to cdsarc.u-strasbg.fr (130.79.128.5) or via \url{http://cdsweb.u-strasbg.fr/cgi-bin/qcat?J/A+A/}

\begin{acknowledgements}
 This work has made use of data from the European Space Agency (ESA) mission Gaia (\url{https://www.cosmos.esa.int/gaia}), processed by the Gaia Data Processing and Analysis Consortium (DPAC, \url{https://www.cosmos.esa.int/web/gaia/dpac/consortium}). Funding for the DPAC has been provided by national institutions, in particular the institutions participating in the Gaia Multilateral Agreement. This work is based on data provided by the Minor Planet Physical Properties Catalogue (MP3C) of the Observatoire de la C\^ote d'Azur. We have received support from France 2030 through the project named Académie Spatiale d'\^Ile-de-France managed by the National Research Agency (ANR-23-CMAS-0041), and from the Centre National d’Etude Spatial (CNES). We would like to thank an anonymous reviewer for the helpful comments and corrections, which improved this manuscript.
\end{acknowledgements}

\bibliographystyle{aa}
\bibliography{refs}

\begin{appendix}
\onecolumn
\section{Additional figures}
\begin{figure*}[h!]
    \centering
    \includegraphics[width=0.9\textwidth]{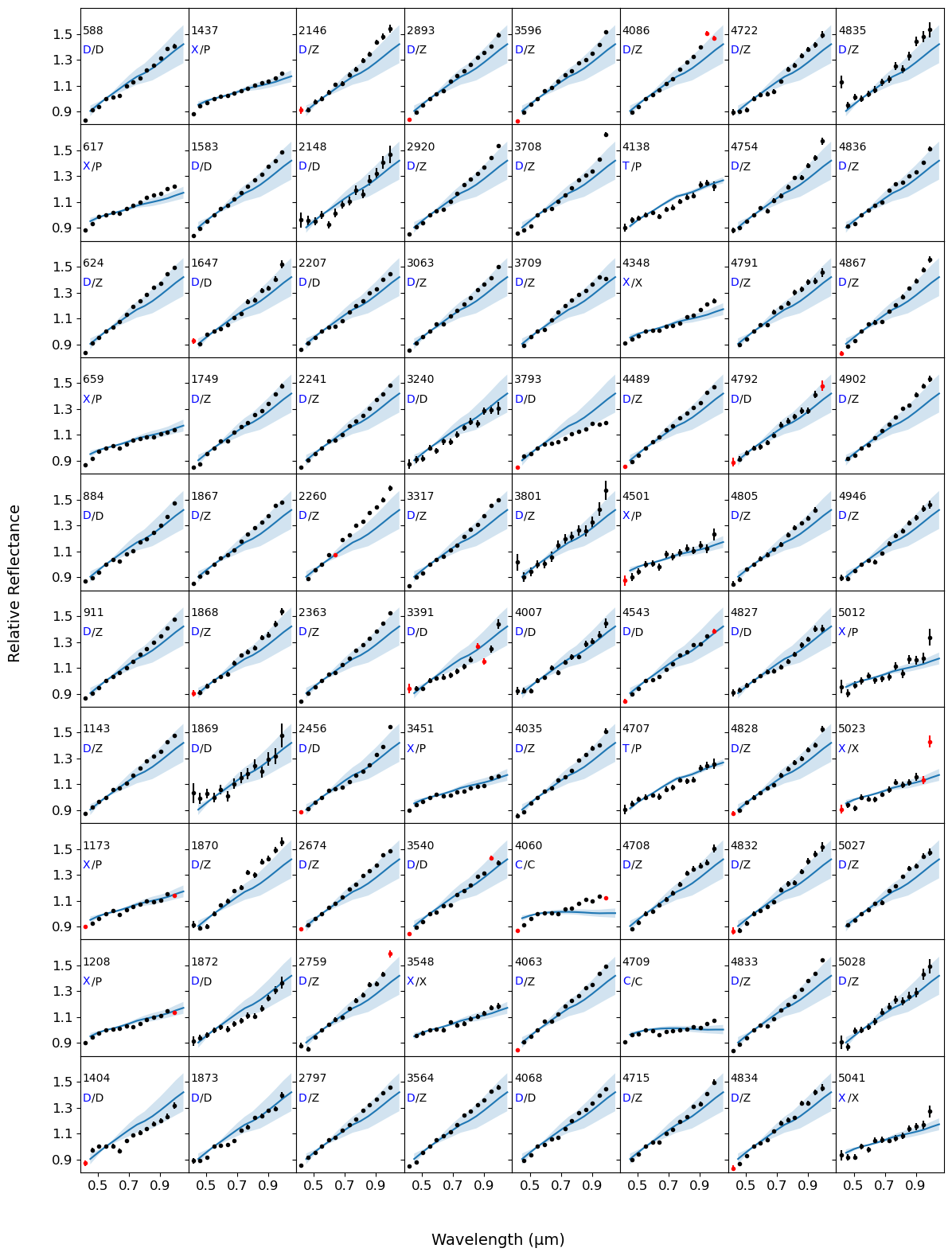}
    \caption{Spectrophotometry of Jupiter Trojans from Gaia DR3 catalog. The red circle indicates data having flag 1 (poor quality). The blue letter indicates the Bus-Demeo taxonomy and the black one the Mahlke's one. The continuous line and blue range represent the best fit mean reflectance spectrum with its uncertainties in the Bus-Demeo taxonomy. }
\label{gaia_1}
\end{figure*}

\begin{figure*}
    \centering
    \includegraphics[width=0.9\textwidth]{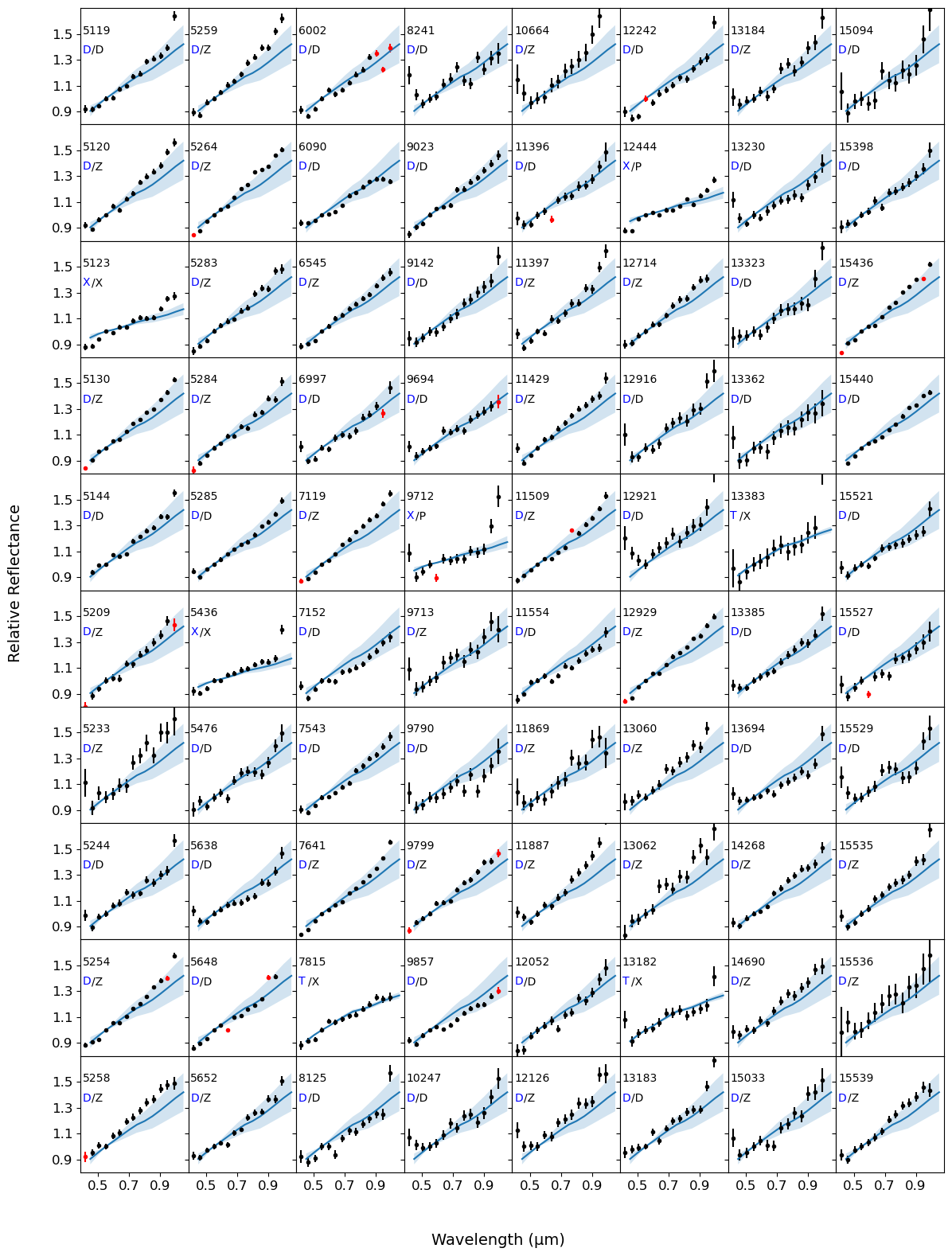}
\caption{Spectrophotometry of Jupiter Trojans from Gaia DR3 catalog. The red circle indicates data having flag 1 (poor quality). The blue letter represents the Bus-Demeo taxonomy and the black one the Mahlke's one. The continuous line and blue range represent the best fit mean reflectance spectrum with its uncertainties in the Bus-Demeo taxonomy.}
\label{gaia_2}
\end{figure*}

\begin{figure*}
    \centering
    \includegraphics[width=0.9\textwidth]{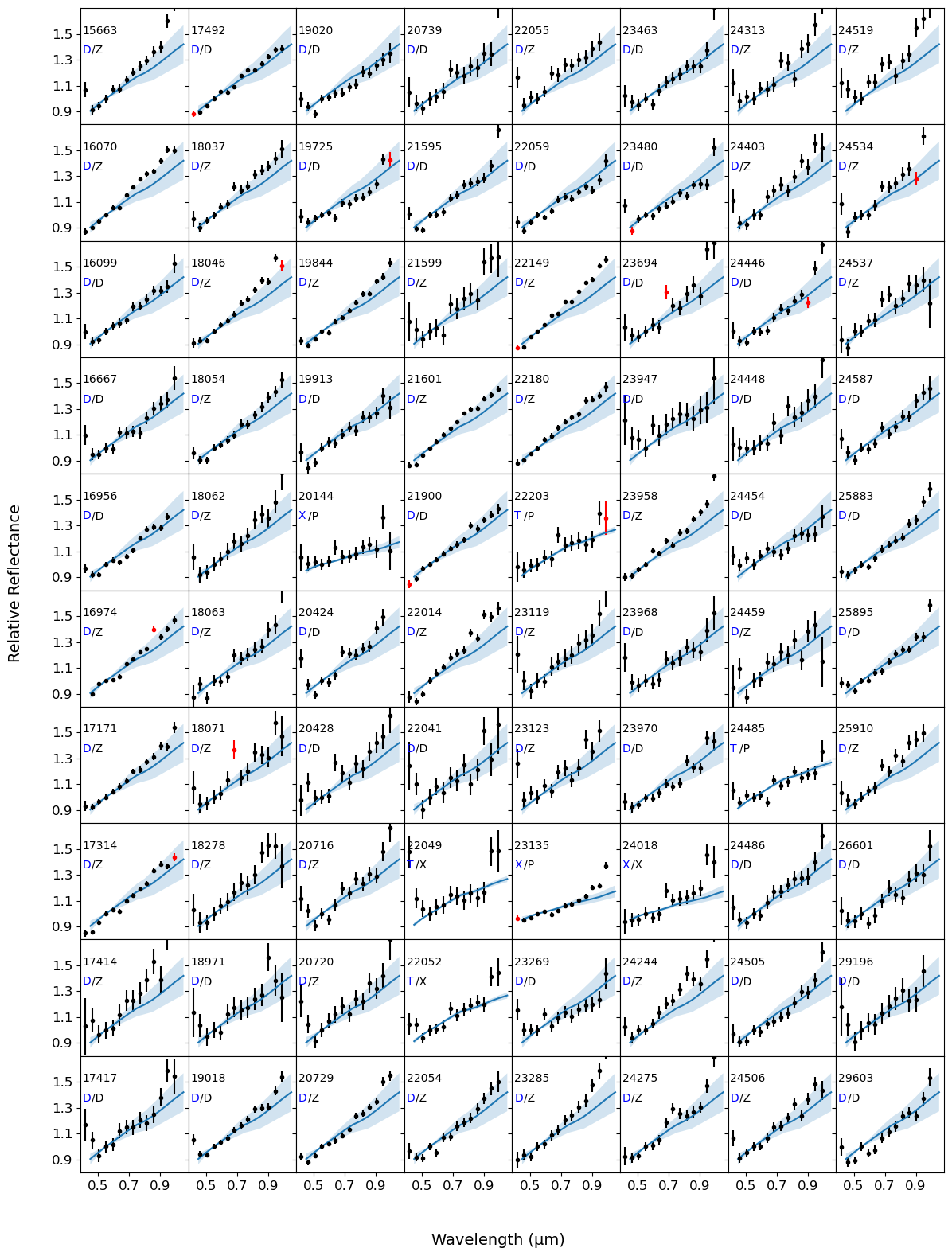}
\caption{Spectrophotometry of Jupiter Trojans from Gaia DR3 catalog. The red circle indicates data having flag 1 (poor quality). The blue letter represents the Bus-Demeo taxonomy and the black one the Mahlke's one. The continuous line and blue range represent the best fit mean reflectance spectrum with its uncertainties in the Bus-Demeo taxonomy.}
\label{gaia_3}
\end{figure*}

\begin{figure*}
    \centering
    \includegraphics[width=0.9\textwidth]{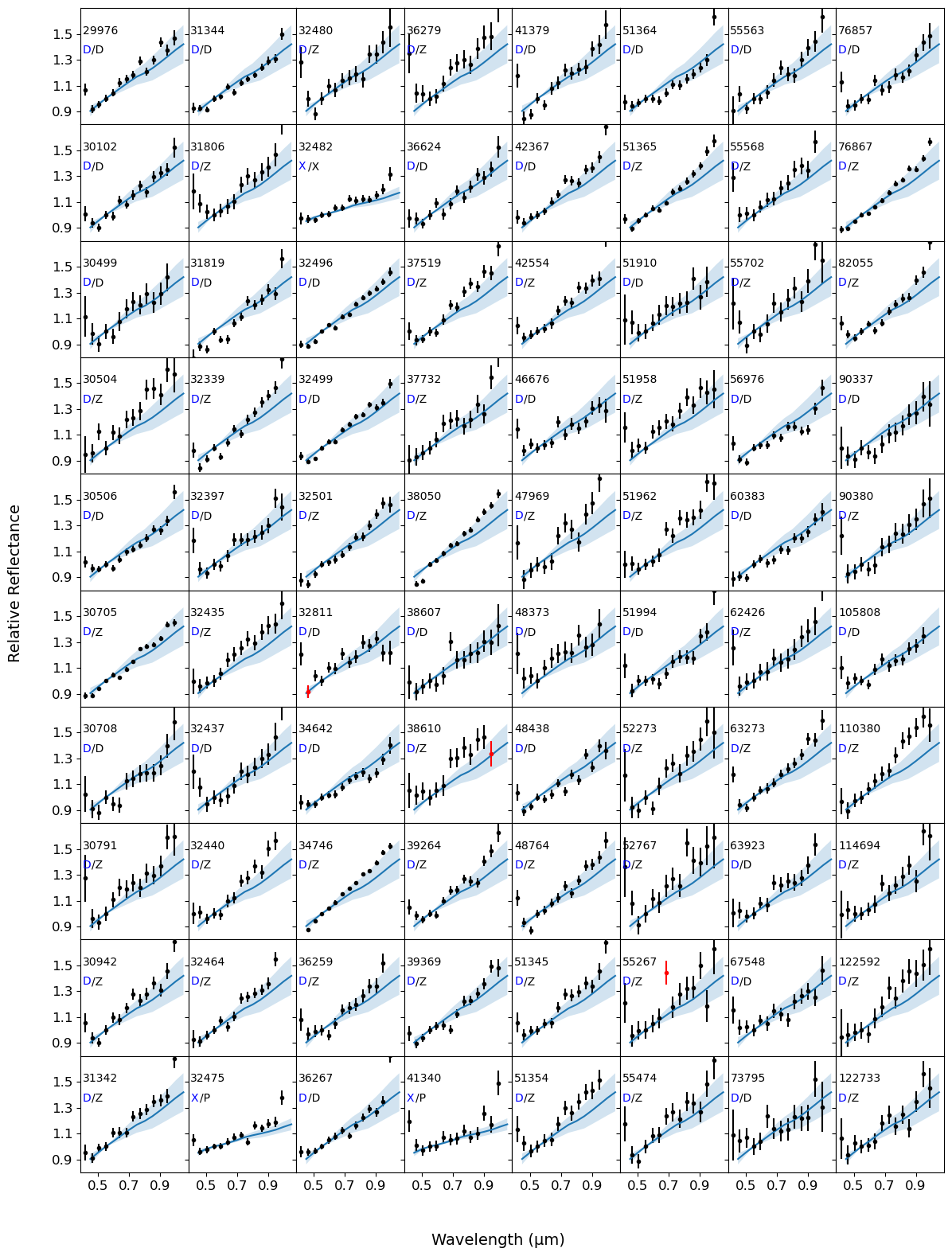}
\caption{Spectrophotometry of Jupiter Trojans from Gaia DR3 catalog. The red circle indicates data having flag 1 (poor quality). The blue letter represents the Bus-Demeo taxonomy and the black one the Mahlke's one. The continuous line and blue range represent the best fit mean reflectance spectrum with its uncertainties in the Bus-Demeo taxonomy.}
\label{gaia_4}
\end{figure*}


\section{Additional Tables}

\begin{center}
\begin{longtable}{c c c c c c c c}
\caption{L4 Trojans investigated. $T_1$ and $T_2$ design the taxonomy using Bus-DeMeo \citep{Demeo_2009} and Mahlke's classification schemes \citep{Mahlke_2022}. The signal to noise ratio (SNR) has been computed for objects having Gaia's spectra. Spectra from the literature are taken from: (1) \citet{galluccio2023}, (2)  \citet{fornasier_2007}, (3) \citet{Roig_2008}, (4) \citet{melita2008physical}, (5) \citet{jewitt}, (6) \citet{Laz}, (7) \citet{Fitz}, (8) \citet{Bendjoya}, (9) \citet{SMASS1}, (10)  \citet{SMASS2}.}\\ \hline\hline
\label{tab:t1}
\textbf{Object} & \textbf{D(km)} & \textbf{S($\%/1000$ \AA)} & \textbf{$p_v$} & \textbf{$T_1$} & \textbf{$T_2$} & \textbf{SNR} & \textbf{References} \\ \hline \endfirsthead
\caption{continued.}\\
\hline \hline
\textbf{Object} & \textbf{D(km)} & \textbf{S($\%/1000$ \AA)} & \textbf{$p_v$} & \textbf{$T_1$} & \textbf{$T_2$} & \textbf{SNR} & \textbf{References} \\ \hline
\endhead
\hline
\endfoot
588 & $131.4 \pm 0.5$ & $8.80 \pm 1.71$ & $0.032 \pm 0.001$ & D & D & 122.4 & 1, 7, 8  \\ 
624 & $196.4 \pm 1.3$ & $10.14 \pm 1.95$ & $0.063 \pm 0.005$ & D & Z & 417.1 & 1, 5  \\ 
659 & $111.7 \pm 1.5$ & $3.36 \pm 1.27$ & $0.034 \pm 0.002$ & X & P & 221.0 & 1, 5  \\ 
911 & $145.0 \pm 1.0$ & $7.74 \pm 1.58$ & $0.045 \pm 0.002$ & D & Z & 230.1 & 1, 6  \\ 
1143 & $116.2 \pm 0.5$ & $11.07 \pm 1.88$ & $0.071 \pm 0.003$ & D & Z & 198.6 & 1, 9  \\ 
1404 & $85.4 \pm 0.9$ & $5.18 \pm 0.77$ & $0.063 \pm 0.003$ & D & D & 70.8 & 1  \\ 
1437 & $127.8 \pm 1.1$ & $3.28 \pm 1.21$ & $0.033 \pm 0.002$ & X & P & 238.3 & 1, 5  \\ 
1583 & $108.7 \pm 0.5$ & $9.55 \pm 1.45$ & $0.057 \pm 0.003$ & D & D & 246.7 & 1, 5  \\ 
1647 & $41.8 \pm 0.7$ & $9.64 \pm 1.48$ & $0.064 \pm 0.013$ & D & D & 72.8 & 1, 5  \\ 
1749 & $66.5 \pm 0.6$ & $9.44 \pm 1.85$ & $0.065 \pm 0.007$ & D & Z & 98.9 & 1, 5, 9  \\ 
1867 & $124.6 \pm 1.0$ & $9.85 \pm 2.37$ & $0.039 \pm 0.001$ & D & Z &  & 5, 7  \\ 
1868 & $70.0 \pm 0.7$ & $8.90 \pm 2.93$ & $0.054 \pm 0.006$ & D & Z & 63.5 & 1, 8  \\ 
1869 & $22.7 \pm 3.4$ & $8.12 \pm 2.16$ & $0.104 \pm 0.031$ & D & D & 25.2 & 1  \\ 
2146 & $50.8 \pm 0.3$ & $9.96 \pm 0.87$ & $0.082 \pm 0.011$ & D & Z & 65.9 & 1  \\ 
2148 & $38.2 \pm 0.3$ & $6.22 \pm 2.37$ & $0.064 \pm 0.003$ & D & D & 31.1 & 1  \\ 
2260 & $77.2 \pm 0.7$ & $12.07 \pm 2.31$ & $0.054 \pm 0.004$ & D & Z & 94.6 & 1, 5, 7  \\ 
2456 & $67.4 \pm 0.5$ & $9.30 \pm 1.64$ & $0.029 \pm 0.002$ & D & D & 132.4 & 1, 5  \\ 
2759 & $55.1 \pm 0.5$ & $8.70 \pm 1.35$ & $0.063 \pm 0.006$ & D & Z & 70.1 & 1, 5  \\ 
2797 & $97.8 \pm 0.6$ & $9.41 \pm 1.15$ & $0.063 \pm 0.003$ & D & Z & 177.4 & 1, 5  \\ 
2920 & $103.9 \pm 1.2$ & $10.34 \pm 1.11$ & $0.043 \pm 0.003$ & D & Z & 165.0 & 1, 5, 9  \\ 
3063 & $109.7 \pm 1.0$ & $8.87 \pm 0.71$ & $0.052 \pm 0.003$ & D & Z & 189.5 & 1, 6  \\ 
3391 & $37.9 \pm 6.8$ & $5.32 \pm 2.17$ & $0.093 \pm 0.036$ & D & D & 48.8 & 1, 5  \\ 
3540 & $69.0 \pm 0.7$ & $8.45 \pm 0.56$ & $0.061 \pm 0.008$ & D & D & 100.4 & 1  \\ 
3548 & $63.8 \pm 0.3$ & $1.67 \pm 2.89$ & $0.053 \pm 0.005$ & X & X & 67.1 & 1, 2  \\ 
3564 & $74.9 \pm 0.4$ & $10.20 \pm 0.52$ & $0.036 \pm 0.001$ & D & Z & 109.2 & 1  \\ 
3596 & $82.9 \pm 0.9$ & $9.74 \pm 0.41$ & $0.044 \pm 0.006$ & D & Z & 140.8 & 1  \\ 
3709 & $69.6 \pm 0.8$ & $10.70 \pm 2.02$ & $0.039 \pm 0.003$ & D & Z & 156.7 & 1, 6, 8  \\ 
3793 & $104.6 \pm 1.4$ & $5.47 \pm 2.44$ & $0.041 \pm 0.004$ & D & D & 138.8 & 1, 5, 6  \\ 
3801 & $34.4 \pm 0.4$ & $11.15 \pm 1.89$ & $0.064 \pm 0.008$ & D & Z & 29.2 & 1  \\ 
4007 & $45.3 \pm 0.3$ & $7.94 \pm 1.02$ & $0.064 \pm 0.009$ & D & D & 52.3 & 1  \\ 
4035 & $67.9 \pm 0.6$ & $9.43 \pm 0.88$ & $0.065 \pm 0.003$ & D & Z & 89.6 & 1, 2, 8  \\ 
4060 & $83.9 \pm 0.5$ & $0.94 \pm 1.38$ & $0.046 \pm 0.003$ & C & C & 132.5 & 1, 6  \\ 
4063 & $97.3 \pm 0.6$ & $8.70 \pm 1.58$ & $0.059 \pm 0.003$ & D & Z & 162.7 & 1, 6, 8  \\ 
4068 & $67.4 \pm 1.2$ & $11.60 \pm 3.33$ & $0.066 \pm 0.008$ & D & D & 128.5 & 1, 6, 8  \\ 
4086 & $85.9 \pm 0.4$ & $10.00 \pm 0.61$ & $0.053 \pm 0.003$ & D & Z & 97.0 & 1  \\ 
4138 & $46.5 \pm 0.4$ & $3.12 \pm 2.16$ & $0.080 \pm 0.005$ & T & P & 49.6 & 1, 8  \\ 
4489 & $78.5 \pm 0.4$ & $9.43 \pm 1.17$ & $0.050 \pm 0.003$ & D & Z & 148.6 & 1, 6  \\ 
4501 & $45.0 \pm 0.5$ & $4.24 \pm 1.43$ & $0.066 \pm 0.008$ & X & P & 36.8 & 1  \\ 
4543 & $62.7 \pm 0.3$ & $8.97 \pm 0.62$ & $0.062 \pm 0.007$ & D & D & 92.2 & 1  \\ 
4833 & $83.1 \pm 0.8$ & $10.73 \pm 1.71$ & $0.049 \pm 0.003$ & D & Z & 147.6 & 1, 6, 8  \\ 
4834 & $73.5 \pm 0.4$ & $10.18 \pm 0.97$ & $0.057 \pm 0.004$ & D & Z & 67.2 & 1, 8  \\ 
4835 & $47.8 \pm 1.1$ & $9.53 \pm 2.40$ & $0.112 \pm 0.032$ & D & Z & 38.1 & 1, 6  \\ 
4836 & $65.3 \pm 0.6$ & $9.13 \pm 2.14$ & $0.065 \pm 0.004$ & D & Z & 120.8 & 1, 8  \\ 
4902 & $52.4 \pm 0.5$ & $9.40 \pm 2.35$ & $0.055 \pm 0.007$ & D & Z & 76.9 & 1, 6  \\ 
4946 & $48.4 \pm 0.4$ & $9.92 \pm 0.90$ & $0.054 \pm 0.005$ & D & Z & 60.8 & 1  \\ 
5012 & $36.8 \pm 0.3$ & $3.37 \pm 1.57$ & $0.083 \pm 0.012$ & X & P & 32.7 & 1  \\ 
5023 & $27.9 \pm 3.5$ & $5.22 \pm 1.22$ & $0.173 \pm 0.093$ & X & X & 44.7 & 1  \\ 
5025 & $46.2 \pm 2.9$ & $13.90 \pm 0.30$ & $0.077 \pm 0.007$ & T & X &  & 8  \\ 
5027 & $60.8 \pm 0.5$ & $10.22 \pm 1.79$ & $0.074 \pm 0.008$ & D & Z & 74.1 & 1, 5  \\ 
5028 & $50.6 \pm 0.3$ & $14.34 \pm 5.07$ & $0.062 \pm 0.006$ & D & Z & 37.1 & 1, 5  \\ 
5041 & $41.6 \pm 0.4$ & $3.71 \pm 1.25$ & $0.059 \pm 0.006$ & X & X & 41.3 & 1  \\ 
5123 & $35.4 \pm 3.0$ & $5.26 \pm 0.87$ & $0.129 \pm 0.015$ & X & X & 58.7 & 1  \\ 
5126 & $57.0 \pm 0.7$ & $0.80 \pm 0.10$ & $0.048 \pm 0.006$ & C & C &  & 8  \\ 
5209 & $47.9 \pm 0.2$ & $9.14 \pm 1.44$ & $0.059 \pm 0.006$ & D & Z & 40.0 & 1  \\ 
5244 & $36.8 \pm 0.5$ & $7.54 \pm 1.37$ & $0.090 \pm 0.011$ & D & D & 40.4 & 1  \\ 
5254 & $76.4 \pm 0.4$ & $10.47 \pm 0.04$ & $0.082 \pm 0.004$ & D & Z & 88.2 & 1, 8  \\ 
5258 & $53.3 \pm 4.4$ & $12.41 \pm 1.39$ & $0.052 \pm 0.014$ & D & Z & 42.1 & 1  \\ 
5259 & $45.2 \pm 0.6$ & $11.66 \pm 1.11$ & $0.074 \pm 0.006$ & D & Z & 50.6 & 1  \\ 
5264 & $69.5 \pm 1.1$ & $10.81 \pm 1.90$ & $0.052 \pm 0.007$ & D & Z & 107.5 & 1, 6, 8  \\ 
5283 & $49.4 \pm 0.4$ & $8.24 \pm 1.06$ & $0.080 \pm 0.008$ & D & Z & 51.1 & 1  \\ 
5284 & $50.2 \pm 0.5$ & $7.45 \pm 0.94$ & $0.070 \pm 0.009$ & D & Z & 58.9 & 1  \\ 
5285 & $52.0 \pm 0.5$ & $7.03 \pm 1.93$ & $0.071 \pm 0.007$ & D & D & 71.5 & 1, 8  \\ 
5436 & $37.6 \pm 0.3$ & $4.74 \pm 1.02$ & $0.091 \pm 0.010$ & X & X & 49.5 & 1  \\ 
5652 & $53.6 \pm 0.4$ & $9.33 \pm 1.12$ & $0.061 \pm 0.008$ & D & Z & 48.7 & 1  \\ 
6090 & $72.3 \pm 0.5$ & $10.40 \pm 2.85$ & $0.047 \pm 0.003$ & D & D & 68.8 & 1, 8  \\ 
6545 & $50.0 \pm 0.4$ & $9.85 \pm 0.92$ & $0.070 \pm 0.008$ & D & Z & 62.5 & 1, 2  \\ 
7119 & $59.3 \pm 0.4$ & $11.80 \pm 0.74$ & $0.046 \pm 0.006$ & D & Z & 78.5 & 1  \\ 
7152 & $39.8 \pm 0.2$ & $4.31 \pm 2.16$ & $0.094 \pm 0.013$ & D & D & 45.0 & 1, 8  \\ 
7214 & $19.9 \pm 0.4$ & $5.12 \pm 0.79$ & $0.112 \pm 0.013$ & X & P &  & 3  \\ 
7543 & $43.5 \pm 0.4$ & $8.70 \pm 0.91$ & $0.055 \pm 0.004$ & D & D & 58.5 & 1  \\ 
7641 & $66.0 \pm 0.9$ & $9.02 \pm 0.54$ & $0.067 \pm 0.005$ & D & Z & 103.0 & 1  \\ 
8125 & $26.6 \pm 0.3$ & $6.17 \pm 1.62$ & $0.122 \pm 0.010$ & D & D & 32.9 & 1  \\ 
8241 & $26.1 \pm 0.6$ & $12.54 \pm 3.72$ & $0.092 \pm 0.016$ & D & D & 26.9 & 1  \\ 
9590 & $ $ & $7.80 \pm 0.87$ & $ $ & D & D &  & 3  \\ 
9694 & $31.7 \pm 0.2$ & $9.03 \pm 1.92$ & $0.101 \pm 0.010$ & D & D & 38.6 & 1  \\ 
9712 & $33.4 \pm 4.3$ & $1.86 \pm 2.02$ & $0.083 \pm 0.026$ & X & P & 26.7 & 1  \\ 
9713 & $18.6 \pm 3.4$ & $12.63 \pm 3.21$ & $0.168 \pm 0.086$ & D & Z & 22.6 & 1  \\ 
9790 & $30.0 \pm 0.8$ & $7.52 \pm 3.09$ & $0.094 \pm 0.016$ & D & D & 22.7 & 1  \\ 
9799 & $65.4 \pm 0.5$ & $9.70 \pm 0.97$ & $0.051 \pm 0.004$ & D & Z & 60.7 & 1  \\ 
9818 & $28.1 \pm 3.2$ & $2.12 \pm 0.22$ & $0.089 \pm 0.021$ & T & P &  & 2  \\ 
9857 & $48.2 \pm 0.4$ & $5.24 \pm 0.90$ & $0.067 \pm 0.008$ & D & D & 60.1 & 1  \\ 
10247 & $27.5 \pm 0.5$ & $11.84 \pm 2.19$ & $0.095 \pm 0.012$ & D & D & 28.1 & 1  \\ 
10664 & $30.8 \pm 0.5$ & $12.15 \pm 2.83$ & $0.068 \pm 0.006$ & D & Z & 20.4 & 1  \\ 
11351 & $32.4 \pm 0.4$ & $10.27 \pm 0.17$ & $0.077 \pm 0.009$ & D & D &  & 2  \\ 
11396 & $37.9 \pm 0.6$ & $7.34 \pm 1.62$ & $0.059 \pm 0.006$ & D & D & 33.4 & 1  \\ 
11397 & $45.1 \pm 0.4$ & $9.93 \pm 1.39$ & $0.079 \pm 0.010$ & D & Z & 39.3 & 1  \\ 
11429 & $37.6 \pm 1.3$ & $10.96 \pm 1.13$ & $0.086 \pm 0.017$ & D & Z & 50.1 & 1  \\ 
12714 & $46.7 \pm 0.6$ & $11.00 \pm 1.20$ & $0.048 \pm 0.005$ & D & Z & 47.6 & 1  \\ 
12916 & $22.3 \pm 1.6$ & $12.14 \pm 2.12$ & $0.098 \pm 0.017$ & D & D & 26.2 & 1  \\ 
12917 & $24.5 \pm 2.3$ & $10.74 \pm 0.15$ & $0.089 \pm 0.027$ & D & Z &  & 2  \\ 
12921 & $32.2 \pm 0.9$ & $9.39 \pm 2.27$ & $0.074 \pm 0.006$ & D & D & 25.3 & 1  \\ 
13060 & $36.1 \pm 2.0$ & $12.64 \pm 1.84$ & $0.071 \pm 0.018$ & D & Z & 31.8 & 1  \\ 
13062 & $28.9 \pm 0.3$ & $13.20 \pm 4.12$ & $0.084 \pm 0.011$ & D & Z & 24.5 & 1, 4  \\ 
13182 & $30.4 \pm 0.3$ & $7.80 \pm 1.92$ & $0.129 \pm 0.024$ & T & X & 27.9 & 1  \\ 
13183 & $40.8 \pm 0.4$ & $10.69 \pm 1.55$ & $0.061 \pm 0.007$ & D & D & 38.4 & 1  \\ 
13184 & $33.9 \pm 0.3$ & $12.04 \pm 2.05$ & $0.068 \pm 0.010$ & D & Z & 27.4 & 1  \\ 
13230 & $23.5 \pm 0.4$ & $6.81 \pm 1.72$ & $0.103 \pm 0.007$ & D & D & 30.3 & 1  \\ 
13323 & $23.4 \pm 0.6$ & $9.24 \pm 2.68$ & $0.129 \pm 0.028$ & D & D & 23.3 & 1, 4  \\ 
13331 & $17.7 \pm 1.5$ & $4.17 \pm 1.22$ & $0.171 \pm 0.033$ & C & C &  & 3  \\ 
13362 & $28.3 \pm 5.2$ & $7.69 \pm 2.83$ & $0.088 \pm 0.051$ & D & D & 18.8 & 1  \\ 
13383 & $23.9 \pm 0.5$ & $6.30 \pm 3.44$ & $0.116 \pm 0.017$ & T & X & 15.7 & 1  \\ 
13385 & $36.3 \pm 0.3$ & $8.64 \pm 1.46$ & $0.067 \pm 0.011$ & D & D & 36.8 & 1  \\ 
13387 & $19.0 \pm 0.8$ & $8.47 \pm 1.71$ & $0.093 \pm 0.022$ & D & D &  & 3  \\ 
13475 & $22.1 \pm 1.9$ & $4.02 \pm 1.11$ & $0.100 \pm 0.038$ & C & C &  & 3  \\ 
13463 & $24.8 \pm 0.4$ & $5.12 \pm 0.12$ & $0.108 \pm 0.014$ & X & P &  & 2  \\ 
13694 & $29.2 \pm 0.3$ & $5.32 \pm 1.42$ & $0.108 \pm 0.010$ & D & D & 36.9 & 1  \\ 
13862 & $25.7 \pm 0.5$ & $1.58 \pm 0.20$ & $0.074 \pm 0.013$ & C & C &  & 2  \\ 
14235 & $27.6 \pm 0.4$ & $9.92 \pm 1.11$ & $0.070 \pm 0.009$ & D & D &  & 3  \\ 
14268 & $49.7 \pm 3.6$ & $12.39 \pm 1.22$ & $0.039 \pm 0.006$ & D & Z & 47.1 & 1  \\ 
14690 & $43.0 \pm 0.5$ & $11.86 \pm 2.21$ & $0.060 \pm 0.007$ & D & Z & 33.9 & 1  \\ 
14707 & $25.9 \pm 0.2$ & $-0.40 \pm 0.24$ & $0.087 \pm 0.015$ & C & C &  & 2  \\ 
15033 & $33.3 \pm 0.5$ & $6.94 \pm 2.18$ & $0.101 \pm 0.020$ & D & Z & 24.9 & 1  \\ 
15094 & $21.0 \pm 0.1$ & $7.59 \pm 3.42$ & $0.103 \pm 0.014$ & D & D & 16.1 & 1  \\ 
15398 & $35.5 \pm 1.1$ & $9.10 \pm 1.58$ & $0.067 \pm 0.016$ & D & D & 35.3 & 1  \\ 
15436 & $88.0 \pm 0.6$ & $10.31 \pm 0.43$ & $0.038 \pm 0.002$ & D & Z & 135.9 & 1  \\ 
15440 & $65.7 \pm 0.7$ & $8.07 \pm 0.61$ & $0.069 \pm 0.006$ & D & D & 92.1 & 1  \\ 
15521 & $26.1 \pm 0.7$ & $8.39 \pm 1.57$ & $0.090 \pm 0.018$ & D & D & 34.5 & 1  \\ 
15527 & $37.1 \pm 0.4$ & $7.24 \pm 2.22$ & $0.059 \pm 0.009$ & D & D & 27.1 & 1  \\ 
15529 & $17.2 \pm 0.8$ & $11.73 \pm 2.31$ & $0.176 \pm 0.042$ & D & D & 24.6 & 1  \\ 
15535 & $40.2 \pm 0.4$ & $11.30 \pm 1.49$ & $0.071 \pm 0.011$ & D & Z & 38.0 & 1  \\ 
15536 & $29.0 \pm 0.7$ & $15.12 \pm 4.99$ & $0.076 \pm 0.009$ & D & Z & 15.1 & 1  \\ 
15539 & $42.2 \pm 2.7$ & $11.20 \pm 2.03$ & $0.082 \pm 0.010$ & D & Z & 37.4 & 1, 4  \\ 
15663 & $35.8 \pm 0.2$ & $11.05 \pm 1.81$ & $0.077 \pm 0.011$ & D & Z & 31.5 & 1  \\ 
16099 & $36.9 \pm 0.2$ & $9.22 \pm 1.76$ & $0.071 \pm 0.009$ & D & D & 31.2 & 1  \\ 
16152 & $16.2 \pm 0.7$ & $4.98 \pm 0.99$ & $0.129 \pm 0.031$ & X & X &  & 3  \\ 
16974 & $57.0 \pm 0.3$ & $10.73 \pm 0.84$ & $0.068 \pm 0.006$ & D & Z & 66.8 & 1  \\ 
18060 & $36.4 \pm 4.0$ & $2.86 \pm 0.10$ & $0.053 \pm 0.009$ & X & P &  & 2  \\ 
18062 & $30.3 \pm 5.6$ & $10.70 \pm 3.49$ & $0.092 \pm 0.060$ & D & Z & 18.2 & 1  \\ 
18063 & $30.6 \pm 0.2$ & $10.21 \pm 2.63$ & $0.080 \pm 0.009$ & D & Z & 21.6 & 1  \\ 
18071 & $20.4 \pm 0.6$ & $9.04 \pm 3.34$ & $0.107 \pm 0.020$ & D & Z & 17.5 & 1  \\ 
19725 & $31.7 \pm 0.5$ & $5.83 \pm 1.62$ & $0.092 \pm 0.019$ & D & D & 34.1 & 1  \\ 
19913 & $25.0 \pm 3.0$ & $8.44 \pm 2.78$ & $0.103 \pm 0.032$ & D & D & 24.8 & 1  \\ 
20144 & $27.5 \pm 0.6$ & $3.57 \pm 2.58$ & $0.071 \pm 0.015$ & X & P & 19.2 & 1  \\ 
20424 & $45.8 \pm 0.5$ & $12.16 \pm 2.01$ & $0.058 \pm 0.008$ & D & D & 27.7 & 1  \\ 
20428 & $27.2 \pm 3.2$ & $10.99 \pm 3.33$ & $0.095 \pm 0.015$ & D & D & 17.4 & 1  \\ 
20716 & $26.7 \pm 0.3$ & $13.59 \pm 2.50$ & $0.055 \pm 0.007$ & D & Z & 22.8 & 1  \\ 
20720 & $34.1 \pm 3.0$ & $9.43 \pm 3.32$ & $0.060 \pm 0.020$ & D & Z & 17.2 & 1  \\ 
20729 & $50.9 \pm 0.9$ & $9.82 \pm 1.14$ & $0.052 \pm 0.010$ & D & Z & 47.9 & 1  \\ 
20738 & $ $ & $9.02 \pm 0.44$ & $ $ & D & D &  & 2  \\ 
20739 & $23.2 \pm 1.1$ & $9.78 \pm 3.31$ & $0.099 \pm 0.019$ & D & D & 17.3 & 1  \\ 
20995 & $26.2 \pm 2.6$ & $11.71 \pm 1.01$ & $0.065 \pm 0.012$ & D & Z &  & 3  \\ 
21370 & $27.8 \pm 0.4$ & $8.73 \pm 1.11$ & $0.058 \pm 0.008$ & D & D &  & 3  \\ 
21595 & $35.4 \pm 0.2$ & $11.05 \pm 1.61$ & $0.077 \pm 0.010$ & D & D & 34.1 & 1  \\ 
21599 & $28.3 \pm 1.0$ & $11.95 \pm 4.07$ & $0.067 \pm 0.013$ & D & Z & 14.5 & 1  \\ 
21601 & $54.7 \pm 0.4$ & $11.82 \pm 0.78$ & $0.063 \pm 0.008$ & D & Z & 72.0 & 1  \\ 
21900 & $50.9 \pm 0.8$ & $9.02 \pm 1.50$ & $0.068 \pm 0.011$ & D & D & 48.8 & 1  \\ 
22014 & $39.5 \pm 0.6$ & $12.44 \pm 1.91$ & $0.103 \pm 0.015$ & D & Z & 38.6 & 1  \\ 
22041 & $21.6 \pm 0.8$ & $9.41 \pm 3.99$ & $0.072 \pm 0.016$ & D & D & 13.7 & 1  \\ 
22049 & $19.1 \pm 0.8$ & $5.85 \pm 3.47$ & $0.111 \pm 0.020$ & T & X & 15.9 & 1  \\ 
22052 & $27.7 \pm 3.2$ & $7.44 \pm 2.46$ & $0.084 \pm 0.032$ & T & X & 22.9 & 1  \\ 
22054 & $32.5 \pm 0.5$ & $9.96 \pm 1.82$ & $0.053 \pm 0.005$ & D & Z & 29.7 & 1  \\ 
22055 & $30.6 \pm 0.5$ & $12.88 \pm 2.56$ & $0.072 \pm 0.009$ & D & Z & 23.0 & 1  \\ 
22059 & $25.2 \pm 0.5$ & $7.97 \pm 1.37$ & $0.104 \pm 0.012$ & D & D & 38.6 & 1  \\ 
22149 & $47.2 \pm 0.3$ & $11.07 \pm 0.73$ & $0.065 \pm 0.010$ & D & Z & 79.8 & 1  \\ 
22203 & $24.4 \pm 0.9$ & $7.59 \pm 3.07$ & $0.074 \pm 0.008$ & T & P & 18.2 & 1  \\ 
22404 & $ $ & $4.16 \pm 0.98$ & $ $ & X & X &  & 3  \\ 
23075 & $33.7 \pm 0.3$ & $12.46 \pm 1.31$ & $0.074 \pm 0.010$ & D & Z &  & 3  \\ 
23119 & $31.1 \pm 0.4$ & $10.45 \pm 3.42$ & $0.059 \pm 0.002$ & D & Z & 16.3 & 1  \\ 
23123 & $25.3 \pm 0.4$ & $12.73 \pm 3.97$ & $0.103 \pm 0.013$ & D & Z & 19.1 & 1  \\ 
23135 & $66.1 \pm 0.6$ & $3.45 \pm 0.80$ & $0.038 \pm 0.005$ & X & P & 66.6 & 1  \\ 
23144 & $17.9 \pm 0.7$ & $6.08 \pm 0.81$ & $0.153 \pm 0.025$ & X & X &  & 3  \\ 
23269 & $22.1 \pm 0.7$ & $6.23 \pm 2.63$ & $0.120 \pm 0.024$ & D & D & 21.4 & 1  \\ 
23285 & $29.7 \pm 2.8$ & $11.60 \pm 2.14$ & $0.087 \pm 0.035$ & D & Z & 27.1 & 1  \\ 
23382 & $24.0 \pm 0.6$ & $9.19 \pm 1.14$ & $0.058 \pm 0.012$ & D & D &  & 3  \\ 
23480 & $26.4 \pm 0.6$ & $7.80 \pm 1.59$ & $0.086 \pm 0.006$ & D & D & 34.3 & 1  \\ 
23706 & $14.1 \pm 0.7$ & $6.78 \pm 1.32$ & $0.097 \pm 0.020$ & D & D &  & 3  \\ 
23939 & $24.5 \pm 1.3$ & $5.42 \pm 0.53$ & $0.089 \pm 0.014$ & X & P &  & 3  \\ 
23947 & $18.9 \pm 1.5$ & $10.37 \pm 4.21$ & $0.078 \pm 0.018$ & D & D & 13.7 & 1  \\ 
23958 & $46.1 \pm 1.1$ & $9.62 \pm 1.14$ & $0.074 \pm 0.012$ & D & Z & 49.3 & 1  \\ 
23963 & $18.7 \pm 0.6$ & $2.57 \pm 1.08$ & $0.127 \pm 0.024$ & C & C &  & 3  \\ 
23968 & $33.0 \pm 0.7$ & $9.01 \pm 2.98$ & $0.067 \pm 0.009$ & D & D & 18.9 & 1  \\ 
23970 & $33.0 \pm 0.4$ & $6.22 \pm 2.35$ & $0.074 \pm 0.012$ & D & D & 29.5 & 1  \\ 
24225 & $14.6 \pm 2.3$ & $4.80 \pm 0.99$ & $0.189 \pm 0.067$ & X & X &  & 3  \\ 
24233 & $22.0 \pm 0.3$ & $6.37 \pm 0.17$ & $0.118 \pm 0.013$ & D & P &  & 2  \\ 
24244 & $34.2 \pm 0.3$ & $14.30 \pm 2.33$ & $0.079 \pm 0.011$ & D & Z & 25.2 & 1  \\ 
24275 & $28.4 \pm 0.8$ & $14.60 \pm 2.07$ & $0.080 \pm 0.011$ & D & Z & 27.7 & 1  \\ 
24313 & $33.9 \pm 0.6$ & $13.36 \pm 4.06$ & $0.067 \pm 0.011$ & D & Z & 18.8 & 1  \\ 
24341 & $16.5 \pm 0.4$ & $-0.26 \pm 0.21$ & $0.154 \pm 0.024$ & C & C &  & 2  \\ 
24380 & $31.1 \pm 0.2$ & $0.34 \pm 0.15$ & $0.069 \pm 0.008$ & D & D &  & 2  \\ 
24390 & $25.4 \pm 1.1$ & $9.53 \pm 0.13$ & $0.069 \pm 0.006$ & D & D &  & 2  \\ 
24403 & $31.1 \pm 0.5$ & $15.07 \pm 3.24$ & $0.068 \pm 0.009$ & D & Z & 23.1 & 1  \\ 
24420 & $22.1 \pm 0.7$ & $1.65 \pm 0.20$ & $0.103 \pm 0.011$ & C & C &  & 2  \\ 
24426 & $14.5 \pm 0.6$ & $4.64 \pm 0.31$ & $0.133 \pm 0.024$ & X & P &  & 2  \\ 
24485 & $26.4 \pm 0.3$ & $6.04 \pm 2.05$ & $0.084 \pm 0.010$ & T & P & 25.6 & 1  \\ 
24486 & $32.6 \pm 0.3$ & $11.40 \pm 2.68$ & $0.068 \pm 0.012$ & D & D & 21.0 & 1  \\ 
24498 & $ $ & $3.12 \pm 1.37$ & $ $ & C & C &  & 3  \\ 
24505 & $29.6 \pm 0.6$ & $6.63 \pm 2.19$ & $0.088 \pm 0.016$ & D & D & 25.1 & 1  \\ 
24508 & $ $ & $5.50 \pm 1.25$ & $ $ & X & X &  & 3  \\ 
24506 & $38.1 \pm 0.5$ & $10.76 \pm 1.90$ & $0.056 \pm 0.007$ & D & Z & 29.1 & 1  \\ 
24519 & $24.7 \pm 0.4$ & $16.32 \pm 3.89$ & $0.075 \pm 0.011$ & D & Z & 20.2 & 1  \\ 
24534 & $29.4 \pm 0.4$ & $13.12 \pm 2.35$ & $0.080 \pm 0.008$ & D & Z & 24.8 & 1  \\ 
24537 & $31.1 \pm 0.4$ & $11.72 \pm 3.08$ & $0.080 \pm 0.010$ & D & Z & 18.5 & 1  \\ 
24539 & $15.5 \pm 1.1$ & $8.54 \pm 1.15$ & $0.128 \pm 0.032$ & D & D &  & 3  \\ 
24882 & $19.1 \pm 2.5$ & $4.43 \pm 1.06$ & $0.133 \pm 0.039$ & X & X &  & 3  \\ 
24587 & $27.2 \pm 0.6$ & $8.41 \pm 2.10$ & $0.082 \pm 0.008$ & D & D & 25.7 & 1  \\ 
25895 & $33.4 \pm 0.5$ & $9.56 \pm 1.34$ & $0.066 \pm 0.006$ & D & D & 41.3 & 1  \\ 
25910 & $23.8 \pm 0.6$ & $10.99 \pm 3.34$ & $0.071 \pm 0.016$ & D & Z & 24.5 & 1  \\ 
26601 & $22.4 \pm 0.4$ & $12.18 \pm 3.84$ & $0.088 \pm 0.019$ & D & D & 18.8 & 1  \\ 
28958 & $22.0 \pm 0.5$ & $-0.04 \pm 0.29$ & $0.065 \pm 0.005$ & C & C &  & 2  \\ 
30102 & $36.2 \pm 0.6$ & $10.18 \pm 1.90$ & $0.078 \pm 0.011$ & D & D & 28.8 & 1  \\ 
31835 & $26.4 \pm 0.5$ & $11.72 \pm 1.43$ & $0.077 \pm 0.009$ & D & Z &  & 3  \\ 
32498 & $21.6 \pm 1.2$ & $11.61 \pm 0.88$ & $0.087 \pm 0.036$ & D & Z &  & 3  \\ 
36259 & $24.6 \pm 2.7$ & $9.63 \pm 2.54$ & $0.088 \pm 0.033$ & D & Z & 23.5 & 1  \\ 
36267 & $37.5 \pm 0.5$ & $7.48 \pm 1.59$ & $0.066 \pm 0.012$ & D & D & 38.1 & 1  \\ 
36279 & $31.0 \pm 0.4$ & $17.54 \pm 4.40$ & $0.065 \pm 0.006$ & D & Z & 17.2 & 1  \\ 
37732 & $21.5 \pm 0.7$ & $9.56 \pm 3.22$ & $0.080 \pm 0.015$ & D & Z & 17.3 & 1  \\ 
38050 & $61.5 \pm 0.3$ & $10.59 \pm 1.02$ & $0.058 \pm 0.007$ & D & Z & 55.8 & 1  \\ 
38606 & $23.2 \pm 0.5$ & $12.07 \pm 1.29$ & $0.073 \pm 0.006$ & D & Z &  & 3  \\ 
38607 & $22.3 \pm 0.4$ & $9.17 \pm 3.49$ & $0.090 \pm 0.012$ & D & D & 16.0 & 1  \\ 
38610 & $27.7 \pm 0.5$ & $19.25 \pm 4.89$ & $0.063 \pm 0.007$ & D & Z & 16.2 & 1  \\ 
38614 & $17.3 \pm 0.9$ & $5.53 \pm 1.05$ & $0.135 \pm 0.032$ & C & C &  & 3  \\ 
38617 & $17.3 \pm 0.9$ & $3.96 \pm 1.00$ & $0.135 \pm 0.032$ & C & C &  & 3  \\ 
38919 & $21.5 \pm 1.8$ & $8.78 \pm 1.31$ & $0.088 \pm 0.042$ & D & D &  & 3  \\ 
39264 & $36.1 \pm 0.3$ & $13.02 \pm 1.68$ & $0.068 \pm 0.009$ & D & Z & 33.3 & 1  \\ 
39285 & $17.5 \pm 0.4$ & $0.25 \pm 0.19$ & $0.061 \pm 0.006$ & C & C &  & 2  \\ 
39369 & $31.9 \pm 0.5$ & $9.20 \pm 1.84$ & $0.060 \pm 0.004$ & D & Z & 30.5 & 1  \\ 
41268 & $15.0 \pm 0.7$ & $10.79 \pm 1.19$ & $0.079 \pm 0.019$ & D & Z &  & 3  \\ 
41340 & $35.0 \pm 3.6$ & $4.91 \pm 2.30$ & $0.058 \pm 0.025$ & X & P & 22.9 & 1  \\ 
41379 & $31.0 \pm 0.4$ & $10.10 \pm 2.59$ & $0.064 \pm 0.009$ & D & D & 22.6 & 1  \\ 
42168 & $19.2 \pm 0.7$ & $3.36 \pm 0.98$ & $0.144 \pm 0.016$ & C & C &  & 3  \\ 
42179 & $15.2 \pm 0.9$ & $4.32 \pm 1.26$ & $0.101 \pm 0.025$ & C & C &  & 3  \\ 
42367 & $33.1 \pm 0.4$ & $13.82 \pm 1.73$ & $0.066 \pm 0.009$ & D & D & 33.7 & 1  \\ 
42554 & $27.4 \pm 2.9$ & $12.02 \pm 1.92$ & $0.059 \pm 0.018$ & D & Z & 28.4 & 1  \\ 
43212 & $19.3 \pm 0.7$ & $1.12 \pm 0.89$ & $0.067 \pm 0.009$ & C & C &  & 3  \\ 
43706 & $14.6 \pm 0.6$ & $5.80 \pm 1.17$ & $0.069 \pm 0.013$ & X & P &  & 3  \\ 
46676 & $18.1 \pm 1.4$ & $7.64 \pm 2.22$ & $0.103 \pm 0.027$ & D & D & 25.6 & 1  \\ 
51378 & $23.2 \pm 1.7$ & $10.49 \pm 0.97$ & $0.042 \pm 0.009$ & D & Z &  & 3  \\ 
53469 & $17.9 \pm 0.7$ & $0.17 \pm 0.30$ & $0.072 \pm 0.016$ & C & C &  & 2  \\ 
53477 & $19.5 \pm 0.8$ & $6.54 \pm 0.87$ & $0.097 \pm 0.021$ & X & P &  & 3  \\ 
55563 & $26.6 \pm 1.3$ & $11.36 \pm 2.62$ & $0.063 \pm 0.013$ & D & D & 21.3 & 1  \\ 
55568 & $27.6 \pm 0.7$ & $10.90 \pm 2.84$ & $0.064 \pm 0.007$ & D & Z & 20.4 & 1  \\ 
57920 & $16.7 \pm 0.5$ & $9.40 \pm 1.31$ & $0.076 \pm 0.019$ & D & D &  & 3  \\ 
58479 & $15.3 \pm 0.9$ & $6.72 \pm 1.28$ & $0.057 \pm 0.012$ & X & P &  & 3  \\ 
60383 & $34.0 \pm 0.4$ & $4.96 \pm 1.81$ & $0.062 \pm 0.008$ & D & D & 29.5 & 1  \\ 
63257 & $14.3 \pm 0.9$ & $4.91 \pm 1.32$ & $0.137 \pm 0.029$ & C & C &  & 3  \\ 
63265 & $18.5 \pm 0.6$ & $11.10 \pm 1.31$ & $0.062 \pm 0.010$ & D & Z &  & 3  \\ 
63273 & $27.6 \pm 0.8$ & $9.63 \pm 1.92$ & $0.070 \pm 0.013$ & D & Z & 29.4 & 1  \\ 
63286 & $16.1 \pm 0.7$ & $7.54 \pm 1.32$ & $0.099 \pm 0.023$ & D & D &  & 3  \\ 
63291 & $12.9 \pm 0.8$ & $6.23 \pm 1.31$ & $0.116 \pm 0.033$ & X & P &  & 3  \\ 
63292 & $18.0 \pm 0.6$ & $4.11 \pm 1.29$ & $0.094 \pm 0.024$ & C & C &  & 3  \\ 
65000 & $24.4 \pm 0.9$ & $9.03 \pm 1.70$ & $0.056 \pm 0.010$ & D & D &  & 3  \\ 
65150 & $ $ & $4.14 \pm 0.20$ & $ $ & X & P &  & 2  \\ 
65194 & $18.6 \pm 0.8$ & $11.31 \pm 2.06$ & $0.081 \pm 0.017$ & D & Z &  & 3  \\ 
65224 & $16.0 \pm 1.2$ & $5.60 \pm 2.01$ & $0.076 \pm 0.017$ & X & P &  & 3  \\ 
65225 & $16.8 \pm 0.2$ & $0.97 \pm 0.34$ & $0.074 \pm 0.008$ & C & C &  & 2  \\ 
89829 & $17.4 \pm 1.6$ & $6.10 \pm 1.07$ & $0.101 \pm 0.023$ & X & X &  & 3  \\ 
89924 & $18.6 \pm 1.6$ & $10.17 \pm 1.37$ & $0.046 \pm 0.013$ & D & D &  & 3  \\ 
90337 & $27.7 \pm 0.6$ & $6.77 \pm 3.55$ & $0.056 \pm 0.009$ & D & D & 15.0 & 1  \\ 
107804 & $18.3 \pm 0.7$ & $14.31 \pm 1.52$ & $0.076 \pm 0.014$ & D & Z &  & 3  \\ 
111819 & $19.3 \pm 0.7$ & $11.69 \pm 1.67$ & $0.043 \pm 0.008$ & D & Z &  & 3  \\ 
114694 & $21.7 \pm 0.6$ & $9.75 \pm 3.25$ & $0.078 \pm 0.014$ & D & Z & 17.4 & 1  \\ 
116954 & $13.6 \pm 1.0$ & $8.05 \pm 1.45$ & $0.050 \pm 0.013$ & D & D &  & 3  \\ 
129602 & $15.7 \pm 1.0$ & $9.60 \pm 1.42$ & $0.113 \pm 0.023$ & D & D &  & 3  \\ 
130190 & $17.2 \pm 0.7$ & $13.13 \pm 1.55$ & $0.114 \pm 0.022$ & D & Z &  & 3  \\ 
137879 & $22.6 \pm 1.0$ & $11.09 \pm 1.18$ & $0.060 \pm 0.012$ & D & Z &  & 3  \\ 
160135 & $ $ & $11.87 \pm 1.75$ & $ $ & D & Z &  & 3  \\ 
160534 & $17.4 \pm 0.6$ & $11.43 \pm 1.55$ & $0.048 \pm 0.008$ & D & Z &  & 3  \\ 
161017 & $13.6 \pm 1.0$ & $3.23 \pm 1.35$ & $0.105 \pm 0.026$ & C & C &  & 3  \\ 
161024 & $ $ & $13.30 \pm 2.42$ & $ $ & D & Z &  & 3  \\ 
163155 & $13.6 \pm 1.7$ & $6.56 \pm 1.17$ & $0.105 \pm 0.037$ & X & P &  & 3  \\ 
162811 & $ $ & $5.41 \pm 1.47$ & $ $ & X & X &  & 3  \\ 
162822 & $17.9 \pm 0.8$ & $12.47 \pm 1.23$ & $0.061 \pm 0.011$ & D & Z &  & 3  \\ 
163135 & $16.2 \pm 0.5$ & $2.76 \pm 0.23$ & $0.060 \pm 0.008$ & X & P &  & 2  \\ 
163216 & $13.0 \pm 0.6$ & $3.60 \pm 0.48$ & $0.094 \pm 0.015$ & X & P &  & 2  \\ 
163702 & $14.8 \pm 1.4$ & $6.10 \pm 1.29$ & $0.127 \pm 0.037$ & X & X &  & 3  \\ 
166148 & $16.9 \pm 1.6$ & $6.85 \pm 2.69$ & $0.039 \pm 0.012$ & X & P &  & 3  \\ 
190294 & $12.0 \pm 0.8$ & $4.33 \pm 1.14$ & $0.102 \pm 0.023$ & X & X &  & 3  \\ 
190446 & $ $ & $12.28 \pm 1.75$ & $ $ & D & Z &  & 3  \\ 
190689 & $ $ & $14.20 \pm 1.86$ & $ $ & D & Z &  & 3  \\ 
191115 & $20.4 \pm 0.7$ & $11.13 \pm 1.66$ & $0.042 \pm 0.011$ & D & Z &  & 3  \\ 
191303 & $ $ & $8.44 \pm 1.54$ & $ $ & D & D &  & 3  \\ 
193535 & $ $ & $4.54 \pm 1.14$ & $ $ & X & X &  & 3  \\ 
192388 & $ $ & $2.76 \pm 0.39$ & $ $ & X & P &  & 2  \\ 
192929 & $13.5 \pm 0.4$ & $-0.53 \pm 0.33$ & $0.119 \pm 0.012$ & C & C &  & 2  \\ 
195084 & $ $ & $6.81 \pm 1.51$ & $ $ & X & X &  & 3  \\ 
195104 & $ $ & $10.54 \pm 1.10$ & $ $ & D & Z &  & 3  \\ 
195218 & $11.8 \pm 0.9$ & $9.06 \pm 1.25$ & $0.126 \pm 0.024$ & D & D &  & 3  \\ 
195490 & $11.2 \pm 0.8$ & $1.77 \pm 1.68$ & $0.107 \pm 0.022$ & C & C &  & 3  \\ 
197624 & $ $ & $4.86 \pm 1.76$ & $ $ & X & X &  & 3  \\ 
202783 & $21.5 \pm 0.6$ & $12.58 \pm 4.50$ & $0.060 \pm 0.014$ & D & Z & 13.3 & 1  \\ 
218070 & $ $ & $5.53 \pm 1.70$ & $ $ & X & X &  & 3  \\ 
219837 & $ $ & $6.98 \pm 1.64$ & $ $ & X & X &  & 3  \\ 
222864 & $ $ & $9.62 \pm 1.54$ & $ $ & D & D &  & 3  \\ 
225623 & $18.5 \pm 0.6$ & $10.80 \pm 1.73$ & $0.068 \pm 0.012$ & D & D &  & 3  \\ 
247019 & $ $ & $13.00 \pm 1.74$ & $ $ & D & Z &  & 3  \\ 
252173 & $ $ & $10.12 \pm 1.31$ & $ $ & D & D &  & 3  \\ 
252683 & $ $ & $0.95 \pm 1.36$ & $ $ & C & C &  & 3  \\ 
252746 & $ $ & $2.90 \pm 1.64$ & $ $ & C & C &  & 3  \\ 
253347 & $ $ & $7.39 \pm 1.68$ & $ $ & X & X &  & 3  \\ 
258628 & $ $ & $10.13 \pm 1.61$ & $ $ & D & D &  & 3  \\ 
258646 & $ $ & $13.27 \pm 1.77$ & $ $ & D & Z &  & 3  \\ \hline
\end{longtable}
\end{center}

\clearpage

\begin{center}
\begin{longtable}{c c c c c c c c}
\caption{L5 Trojans investigated. $T_1$ and $T_2$ design the taxonomy using Bus-DeMeo \citep{Demeo_2009} and Mahlke's classification schemes \citep{Mahlke_2022}. The signal to noise ratio (SNR) has been computed for objects having Gaia's spectra. Spectra from the literature are taken from: (1) \citet{galluccio2023}, (2)  \citet{fornasier_2007}, (3) \citet{Roig_2008}, (4) \citet{melita2008physical}, (5) \citet{jewitt}, (6) \citet{Laz}, (7) \citet{Fitz}, (8) \citet{Bendjoya}, (9) \citet{SMASS1}, (10)  \citet{SMASS2}.}\\
\hline\hline
\label{tab:t2}
\textbf{Object} & \textbf{D(km)} & \textbf{S($\%/1000$ \AA)} & \textbf{$p_v$} & \textbf{$T_1$} & \textbf{$T_2$} & \textbf{SNR} & \textbf{References} \\ \hline \endfirsthead
\caption{continued.}\\
\hline\hline
\textbf{Object} & \textbf{D(km)} & \textbf{S($\%/1000$ \AA)} & \textbf{$p_v$} & \textbf{$T_1$} & \textbf{$T_2$} & \textbf{SNR} & \textbf{References} \\ \hline
\endhead
\hline
\endfoot
617 & $140.7 \pm 0.7$ & $4.50 \pm 0.25$ & $0.042 \pm 0.001$ & X & P & 215.9 & 1  \\ 
884 & $102.5 \pm 0.5$ & $7.02 \pm 0.46$ & $0.044 \pm 0.002$ & D & D & 120.3 & 1  \\ 
1172 & $125.9 \pm 0.7$ & $10.04 \pm 0.09$ & $0.043 \pm 0.002$ & D & Z &  & 2  \\ 
1173 & $110.5 \pm 0.7$ & $2.76 \pm 0.54$ & $0.033 \pm 0.002$ & X & P & 134.1 & 1, 4  \\ 
1208 & $104.2 \pm 0.7$ & $2.13 \pm 0.40$ & $0.039 \pm 0.002$ & X & P & 130.2 & 1  \\ 
1867 & $124.6 \pm 1.0$ & $10.17 \pm 0.20$ & $0.039 \pm 0.001$ & D & Z & 292.8 & 1  \\ 
1870 & $46.6 \pm 0.5$ & $13.31 \pm 0.92$ & $0.051 \pm 0.005$ & D & Z & 61.0 & 1  \\ 
1871 & $27.7 \pm 0.4$ & $7.13 \pm 0.26$ & $0.081 \pm 0.010$ & D & D &  & 2  \\ 
1872 & $33.3 \pm 0.4$ & $4.67 \pm 1.21$ & $0.074 \pm 0.010$ & D & D & 44.0 & 1  \\ 
1873 & $49.3 \pm 0.6$ & $7.26 \pm 0.76$ & $0.051 \pm 0.004$ & D & D & 70.1 & 1  \\ 
2207 & $97.0 \pm 0.5$ & $8.83 \pm 0.26$ & $0.054 \pm 0.003$ & D & D & 214.2 & 1  \\ 
2223 & $78.3 \pm 0.6$ & $10.20 \pm 0.14$ & $0.038 \pm 0.002$ & D & D &  & 2  \\ 
2241 & $112.6 \pm 1.0$ & $9.26 \pm 0.31$ & $0.046 \pm 0.002$ & D & Z & 185.4 & 1  \\ 
2357 & $97.7 \pm 0.9$ & $9.92 \pm 0.19$ & $0.049 \pm 0.003$ & D & Z &  & 2  \\ 
2363 & $89.3 \pm 0.7$ & $10.42 \pm 0.24$ & $0.069 \pm 0.003$ & D & Z & 239.3 & 1  \\ 
2674 & $74.4 \pm 0.3$ & $10.40 \pm 0.38$ & $0.048 \pm 0.003$ & D & Z & 153.7 & 1  \\ 
2893 & $87.8 \pm 0.8$ & $9.94 \pm 0.44$ & $0.039 \pm 0.002$ & D & Z & 131.4 & 1  \\ 
2895 & $56.6 \pm 0.2$ & $0.10 \pm 0.60$ & $0.063 \pm 0.011$ & C & C &  & 8  \\ 
3240 & $51.7 \pm 0.2$ & $7.55 \pm 1.34$ & $0.059 \pm 0.009$ & D & D & 38.9 & 1  \\ 
3317 & $116.8 \pm 0.7$ & $9.15 \pm 2.01$ & $0.059 \pm 0.003$ & D & Z & 153.5 & 1, 8, 10  \\ 
3451 & $116.8 \pm 0.7$ & $1.81 \pm 0.30$ & $0.059 \pm 0.003$ & X & P & 318.1 & 1, 10  \\ 
3708 & $77.6 \pm 0.7$ & $9.53 \pm 0.90$ & $0.058 \pm 0.003$ & D & Z & 99.1 & 1, 8  \\ 
4348 & $82.1 \pm 0.6$ & $3.55 \pm 1.08$ & $0.035 \pm 0.004$ & X & X & 127.8 & 1, 8  \\ 
4707 & $37.8 \pm 0.4$ & $5.71 \pm 1.18$ & $0.085 \pm 0.014$ & T & P & 44.2 & 1  \\ 
4708 & $55.0 \pm 0.5$ & $10.24 \pm 0.93$ & $0.064 \pm 0.005$ & D & Z & 61.4 & 1  \\ 
4709 & $88.8 \pm 0.7$ & $0.19 \pm 0.25$ & $0.078 \pm 0.006$ & C & C & 204.2 & 1  \\ 
4715 & $63.0 \pm 0.4$ & $11.46 \pm 3.65$ & $0.058 \pm 0.007$ & D & Z & 94.6 & 1, 8  \\ 
4722 & $50.7 \pm 0.3$ & $9.11 \pm 0.88$ & $0.069 \pm 0.009$ & D & Z & 62.8 & 1  \\ 
4754 & $52.5 \pm 0.6$ & $10.90 \pm 2.46$ & $0.060 \pm 0.006$ & D & Z & 68.8 & 1, 4  \\ 
4791 & $49.3 \pm 0.5$ & $10.34 \pm 0.90$ & $0.073 \pm 0.004$ & D & Z & 62.7 & 1  \\ 
4792 & $50.0 \pm 0.5$ & $10.17 \pm 1.12$ & $0.069 \pm 0.007$ & D & D & 51.1 & 1  \\ 
4805 & $56.9 \pm 0.7$ & $9.73 \pm 0.86$ & $0.062 \pm 0.009$ & D & Z & 66.1 & 1  \\ 
4827 & $42.8 \pm 0.2$ & $6.32 \pm 0.93$ & $0.067 \pm 0.006$ & D & D & 57.9 & 1  \\ 
4828 & $46.4 \pm 0.3$ & $9.85 \pm 0.76$ & $0.061 \pm 0.006$ & D & Z & 76.1 & 1  \\ 
4829 & $32.0 \pm 0.3$ & $5.03 \pm 0.19$ & $0.069 \pm 0.007$ & T & P &  & 2  \\ 
4832 & $51.6 \pm 0.4$ & $10.52 \pm 0.97$ & $0.072 \pm 0.006$ & D & Z & 58.2 & 1  \\ 
4867 & $59.0 \pm 0.6$ & $8.39 \pm 0.70$ & $0.063 \pm 0.006$ & D & Z & 81.7 & 1  \\ 
5119 & $50.6 \pm 0.4$ & $9.60 \pm 1.05$ & $0.062 \pm 0.006$ & D & D & 52.1 & 1  \\ 
5120 & $48.8 \pm 0.4$ & $10.32 \pm 0.94$ & $0.118 \pm 0.019$ & D & Z & 59.2 & 1  \\ 
5130 & $60.4 \pm 0.7$ & $10.07 \pm 0.59$ & $0.067 \pm 0.007$ & D & Z & 98.9 & 1  \\ 
5144 & $85.1 \pm 0.4$ & $8.85 \pm 0.80$ & $0.050 \pm 0.002$ & D & D & 70.4 & 1  \\ 
5233 & $29.0 \pm 0.3$ & $14.53 \pm 2.79$ & $0.061 \pm 0.010$ & D & Z & 21.1 & 1  \\ 
5476 & $34.5 \pm 0.4$ & $9.78 \pm 1.83$ & $0.103 \pm 0.015$ & D & D & 30.8 & 1  \\ 
5511 & $39.4 \pm 0.3$ & $10.84 \pm 0.15$ & $0.098 \pm 0.011$ & D & D &  & 2  \\ 
5638 & $41.0 \pm 0.4$ & $5.01 \pm 1.23$ & $0.060 \pm 0.007$ & D & D & 43.8 & 1  \\ 
5648 & $59.5 \pm 1.3$ & $10.64 \pm 4.50$ & $0.070 \pm 0.011$ & D & D & 72.9 & 1, 6  \\ 
6002 & $39.9 \pm 0.4$ & $6.96 \pm 1.03$ & $0.076 \pm 0.008$ & D & D & 52.9 & 1  \\ 
6997 & $37.2 \pm 0.2$ & $6.53 \pm 1.36$ & $0.090 \pm 0.009$ & D & D & 40.0 & 1  \\ 
6998 & $27.6 \pm 0.3$ & $11.30 \pm 0.24$ & $0.056 \pm 0.007$ & D & Z &  & 2  \\ 
7352 & $47.5 \pm 0.6$ & $10.30 \pm 0.20$ & $0.091 \pm 0.010$ & D & D &  & 8  \\ 
7815 & $42.5 \pm 1.0$ & $4.97 \pm 1.10$ & $0.089 \pm 0.008$ & T & X & 48.5 & 1  \\ 
9023 & $48.2 \pm 0.5$ & $9.21 \pm 1.03$ & $0.057 \pm 0.007$ & D & D & 52.5 & 1  \\ 
9030 & $33.0 \pm 0.5$ & $10.36 \pm 0.25$ & $0.063 \pm 0.006$ & D & D &  & 2  \\ 
9142 & $42.3 \pm 0.5$ & $9.93 \pm 1.93$ & $0.066 \pm 0.012$ & D & D & 28.8 & 1  \\ 
9430 & $27.1 \pm 0.3$ & $10.02 \pm 0.40$ & $0.083 \pm 0.009$ & D & D &  & 2  \\ 
11089 & $37.1 \pm 0.4$ & $4.53 \pm 0.17$ & $0.076 \pm 0.011$ & X & P &  & 2  \\ 
11488 & $22.1 \pm 0.4$ & $6.67 \pm 0.15$ & $0.119 \pm 0.011$ & X & P &  & 2  \\ 
11509 & $50.1 \pm 0.7$ & $6.97 \pm 0.94$ & $0.062 \pm 0.009$ & D & Z & 74.2 & 1  \\ 
11552 & $49.8 \pm 0.4$ & $9.57 \pm 0.09$ & $0.065 \pm 0.006$ & D & Z &  & 4  \\ 
11554 & $41.6 \pm 0.5$ & $3.94 \pm 1.19$ & $0.065 \pm 0.003$ & D & D & 47.3 & 1  \\ 
11663 & $31.3 \pm 4.9$ & $6.91 \pm 0.20$ & $0.086 \pm 0.063$ & D & D &  & 2  \\ 
11869 & $26.1 \pm 0.3$ & $9.21 \pm 3.63$ & $0.053 \pm 0.008$ & D & Z & 20.1 & 1  \\ 
11887 & $30.8 \pm 0.4$ & $10.69 \pm 1.45$ & $0.090 \pm 0.007$ & D & Z & 40.0 & 1  \\ 
12052 & $38.8 \pm 0.6$ & $6.88 \pm 2.33$ & $0.071 \pm 0.009$ & D & D & 33.0 & 1  \\ 
12126 & $52.0 \pm 0.4$ & $11.23 \pm 1.89$ & $0.052 \pm 0.007$ & D & Z & 30.4 & 1  \\ 
12242 & $36.9 \pm 0.4$ & $10.41 \pm 1.89$ & $0.079 \pm 0.012$ & D & D & 38.7 & 1  \\ 
12444 & $62.1 \pm 0.4$ & $2.17 \pm 1.01$ & $0.040 \pm 0.002$ & X & P & 67.5 & 1  \\ 
12929 & $52.1 \pm 0.4$ & $10.02 \pm 0.67$ & $0.068 \pm 0.008$ & D & Z & 84.5 & 1  \\ 
15502 & $53.1 \pm 0.1$ & $11.06 \pm 0.21$ & $0.055 \pm 0.009$ & D & D &  & 2  \\ 
15977 & $42.2 \pm 0.5$ & $8.66 \pm 0.14$ & $0.069 \pm 0.007$ & D & D &  & 2  \\ 
16070 & $63.9 \pm 0.9$ & $12.44 \pm 0.90$ & $0.050 \pm 0.005$ & D & Z & 62.6 & 1  \\ 
16560 & $43.6 \pm 0.4$ & $3.43 \pm 0.04$ & $0.039 \pm 0.004$ & C & C &  & 4  \\ 
16667 & $35.4 \pm 0.4$ & $6.88 \pm 2.23$ & $0.062 \pm 0.008$ & D & D & 24.3 & 1  \\ 
16956 & $35.6 \pm 0.6$ & $7.92 \pm 1.14$ & $0.096 \pm 0.018$ & D & D & 48.1 & 1  \\ 
17171 & $41.6 \pm 0.4$ & $10.19 \pm 1.22$ & $0.077 \pm 0.007$ & D & Z & 46.3 & 1  \\ 
17314 & $35.8 \pm 0.4$ & $8.41 \pm 0.91$ & $0.072 \pm 0.012$ & D & Z & 60.6 & 1  \\ 
17414 & $21.4 \pm 0.3$ & $14.39 \pm 4.07$ & $0.063 \pm 0.007$ & D & Z & 14.4 & 1  \\ 
17416 & $17.6 \pm 0.5$ & $10.80 \pm 0.35$ & $0.068 \pm 0.008$ & D & D &  & 2  \\ 
17417 & $27.6 \pm 0.8$ & $9.63 \pm 2.90$ & $0.049 \pm 0.010$ & D & D & 19.4 & 1  \\ 
17419 & $34.8 \pm 0.5$ & $3.23 \pm 0.88$ & $0.067 \pm 0.010$ & C & C &  & 3  \\ 
17420 & $18.4 \pm 1.0$ & $7.50 \pm 1.05$ & $0.100 \pm 0.017$ & D & D &  & 3  \\ 
17492 & $54.5 \pm 0.6$ & $9.57 \pm 0.89$ & $0.065 \pm 0.007$ & D & D & 64.3 & 1  \\ 
18037 & $25.0 \pm 0.8$ & $10.19 \pm 1.72$ & $0.082 \pm 0.009$ & D & Z & 33.2 & 1  \\ 
18046 & $41.5 \pm 0.5$ & $11.60 \pm 1.18$ & $0.078 \pm 0.009$ & D & Z & 48.1 & 1  \\ 
18054 & $36.4 \pm 0.4$ & $9.14 \pm 1.62$ & $0.071 \pm 0.005$ & D & Z & 34.4 & 1  \\ 
18137 & $33.9 \pm 0.4$ & $8.66 \pm 0.14$ & $0.067 \pm 0.007$ & D & D &  & 2  \\ 
18268 & $20.2 \pm 0.5$ & $14.27 \pm 0.21$ & $0.060 \pm 0.008$ & D & D &  & 2  \\ 
18278 & $29.4 \pm 0.3$ & $11.24 \pm 3.51$ & $0.053 \pm 0.007$ & D & Z & 16.5 & 1  \\ 
18493 & $32.7 \pm 0.3$ & $5.17 \pm 0.30$ & $0.091 \pm 0.011$ & D & D &  & 2  \\ 
18940 & $21.6 \pm 0.6$ & $7.29 \pm 1.00$ & $0.105 \pm 0.008$ & D & D &  & 2, 4  \\ 
18971 & $24.8 \pm 0.5$ & $9.83 \pm 3.90$ & $0.060 \pm 0.012$ & D & D & 14.3 & 1  \\ 
19018 & $31.0 \pm 0.3$ & $9.69 \pm 1.27$ & $0.077 \pm 0.012$ & D & D & 44.2 & 1  \\ 
19020 & $42.9 \pm 0.3$ & $5.15 \pm 1.88$ & $0.070 \pm 0.010$ & D & D & 27.9 & 1  \\ 
19844 & $36.5 \pm 0.3$ & $10.68 \pm 0.97$ & $0.059 \pm 0.007$ & D & Z & 57.7 & 1  \\ 
22180 & $39.8 \pm 2.8$ & $10.70 \pm 1.04$ & $0.102 \pm 0.022$ & D & Z & 54.0 & 1  \\ 
23463 & $25.5 \pm 0.4$ & $8.90 \pm 2.58$ & $0.077 \pm 0.009$ & D & D & 23.2 & 1  \\ 
23549 & $17.4 \pm 0.5$ & $8.49 \pm 0.35$ & $0.139 \pm 0.018$ & D & D &  & 2  \\ 
23694 & $32.4 \pm 1.5$ & $9.13 \pm 2.79$ & $0.051 \pm 0.017$ & D & D & 20.5 & 1  \\ 
24018 & $23.9 \pm 0.3$ & $6.58 \pm 2.75$ & $0.082 \pm 0.009$ & X & X & 20.1 & 1  \\ 
24022 & $21.0 \pm 0.5$ & $11.46 \pm 0.22$ & $0.070 \pm 0.014$ & D & Z &  & 4  \\ 
24444 & $23.5 \pm 0.5$ & $6.05 \pm 0.15$ & $0.106 \pm 0.014$ & X & X &  & 2  \\ 
24446 & $32.0 \pm 0.6$ & $9.22 \pm 1.84$ & $0.085 \pm 0.009$ & D & D & 29.4 & 1  \\ 
24448 & $25.4 \pm 0.8$ & $5.06 \pm 4.88$ & $0.076 \pm 0.018$ & D & D & 16.2 & 1  \\ 
24452 & $18.0 \pm 0.8$ & $7.42 \pm 0.18$ & $0.192 \pm 0.025$ & D & D &  & 2  \\ 
24454 & $27.8 \pm 0.4$ & $4.28 \pm 2.38$ & $0.055 \pm 0.008$ & D & D & 22.8 & 1  \\ 
24459 & $27.4 \pm 0.4$ & $10.73 \pm 3.45$ & $0.061 \pm 0.009$ & D & Z & 16.2 & 1  \\ 
24467 & $20.9 \pm 0.4$ & $12.45 \pm 0.17$ & $0.089 \pm 0.009$ & D & D &  & 2  \\ 
25347 & $20.9 \pm 0.4$ & $10.11 \pm 0.33$ & $0.089 \pm 0.009$ & D & Z &  & 2  \\ 
25883 & $29.3 \pm 0.2$ & $9.70 \pm 1.56$ & $0.064 \pm 0.008$ & D & D & 36.0 & 1  \\ 
29196 & $16.1 \pm 1.9$ & $10.77 \pm 4.27$ & $0.142 \pm 0.052$ & D & D & 13.4 & 1  \\ 
29314 & $21.4 \pm 0.8$ & $4.29 \pm 1.13$ & $0.140 \pm 0.039$ & X & X &  & 3  \\ 
29603 & $31.4 \pm 0.4$ & $8.64 \pm 1.89$ & $0.070 \pm 0.010$ & D & D & 28.6 & 1  \\ 
29976 & $33.8 \pm 0.5$ & $10.89 \pm 2.06$ & $0.081 \pm 0.009$ & D & D & 35.7 & 1  \\ 
30499 & $21.2 \pm 0.6$ & $12.50 \pm 3.61$ & $0.075 \pm 0.016$ & D & D & 15.3 & 1  \\ 
30504 & $27.6 \pm 0.6$ & $12.43 \pm 3.40$ & $0.066 \pm 0.005$ & D & Z & 17.6 & 1  \\ 
30505 & $24.5 \pm 0.4$ & $12.57 \pm 1.01$ & $0.083 \pm 0.009$ & D & Z &  & 3  \\ 
30506 & $35.2 \pm 0.5$ & $7.04 \pm 1.50$ & $0.067 \pm 0.011$ & D & D & 37.9 & 1  \\ 
30698 & $18.3 \pm 0.9$ & $9.08 \pm 0.32$ & $0.101 \pm 0.014$ & D & D &  & 2  \\ 
30705 & $45.7 \pm 0.6$ & $9.76 \pm 0.82$ & $0.073 \pm 0.008$ & D & Z & 66.8 & 1  \\ 
30708 & $25.7 \pm 0.5$ & $10.13 \pm 3.25$ & $0.068 \pm 0.005$ & D & D & 16.7 & 1  \\ 
30791 & $20.3 \pm 2.0$ & $9.75 \pm 3.35$ & $0.108 \pm 0.039$ & D & Z & 17.4 & 1  \\ 
30942 & $30.6 \pm 0.4$ & $11.79 \pm 2.21$ & $0.064 \pm 0.009$ & D & Z & 26.0 & 1  \\ 
31342 & $47.4 \pm 0.3$ & $10.63 \pm 2.09$ & $0.057 \pm 0.008$ & D & Z & 28.0 & 1  \\ 
31344 & $40.5 \pm 0.5$ & $7.23 \pm 1.37$ & $0.047 \pm 0.003$ & D & D & 40.7 & 1  \\ 
31806 & $23.4 \pm 0.2$ & $13.31 \pm 3.09$ & $0.083 \pm 0.011$ & D & Z & 18.8 & 1  \\ 
31814 & $18.1 \pm 0.8$ & $12.17 \pm 1.37$ & $0.085 \pm 0.016$ & D & Z &  & 3  \\ 
31819 & $28.7 \pm 0.6$ & $11.12 \pm 1.79$ & $0.065 \pm 0.008$ & D & D & 29.9 & 1  \\ 
31820 & $16.8 \pm 0.5$ & $7.53 \pm 0.29$ & $0.094 \pm 0.015$ & D & D &  & 2  \\ 
31821 & $19.9 \pm 0.2$ & $10.57 \pm 0.32$ & $0.078 \pm 0.006$ & D & D &  & 2  \\ 
32339 & $25.3 \pm 0.8$ & $11.27 \pm 1.72$ & $0.060 \pm 0.006$ & D & Z & 32.4 & 1  \\ 
32397 & $29.8 \pm 0.5$ & $11.09 \pm 2.45$ & $0.064 \pm 0.007$ & D & D & 22.8 & 1  \\ 
32430 & $13.4 \pm 0.5$ & $8.11 \pm 0.56$ & $0.158 \pm 0.007$ & D & D &  & 2  \\ 
32435 & $31.0 \pm 0.2$ & $15.09 \pm 3.69$ & $0.087 \pm 0.012$ & D & Z & 20.9 & 1  \\ 
32437 & $28.7 \pm 0.2$ & $10.39 \pm 3.27$ & $0.078 \pm 0.011$ & D & D & 16.9 & 1  \\ 
32440 & $29.0 \pm 0.4$ & $14.20 \pm 2.43$ & $0.063 \pm 0.005$ & D & Z & 24.4 & 1  \\ 
32464 & $29.3 \pm 0.3$ & $11.62 \pm 1.99$ & $0.061 \pm 0.005$ & D & Z & 29.5 & 1  \\ 
32467 & $20.1 \pm 0.4$ & $12.88 \pm 0.09$ & $0.091 \pm 0.016$ & D & Z &  & 4  \\ 
32475 & $40.0 \pm 0.3$ & $5.62 \pm 1.86$ & $0.076 \pm 0.011$ & X & P & 36.4 & 1  \\ 
32480 & $29.1 \pm 0.2$ & $8.44 \pm 3.88$ & $0.076 \pm 0.006$ & D & Z & 18.8 & 1  \\ 
32482 & $27.8 \pm 0.4$ & $5.92 \pm 1.45$ & $0.109 \pm 0.018$ & X & X & 36.8 & 1  \\ 
32496 & $48.2 \pm 0.9$ & $8.56 \pm 0.94$ & $0.067 \pm 0.013$ & D & D & 57.5 & 1  \\ 
32499 & $38.9 \pm 0.3$ & $10.87 \pm 1.01$ & $0.081 \pm 0.008$ & D & D & 55.7 & 1  \\ 
32501 & $35.6 \pm 0.4$ & $8.98 \pm 1.57$ & $0.062 \pm 0.005$ & D & Z & 34.7 & 1  \\ 
32615 & $35.4 \pm 0.3$ & $9.39 \pm 0.16$ & $0.066 \pm 0.004$ & D & D &  & 2  \\ 
32794 & $13.9 \pm 0.8$ & $6.59 \pm 0.38$ & $0.081 \pm 0.007$ & D & D &  & 2  \\ 
32811 & $28.2 \pm 0.4$ & $7.96 \pm 2.36$ & $0.080 \pm 0.014$ & D & D & 23.8 & 1  \\ 
34298 & $17.2 \pm 1.9$ & $8.62 \pm 1.33$ & $0.086 \pm 0.020$ & D & D &  & 3  \\ 
34642 & $33.7 \pm 0.3$ & $7.75 \pm 1.64$ & $0.097 \pm 0.009$ & D & D & 32.6 & 1  \\ 
34746 & $61.1 \pm 0.7$ & $11.23 \pm 0.71$ & $0.061 \pm 0.004$ & D & Z & 81.6 & 1  \\ 
34785 & $28.4 \pm 0.3$ & $1.73 \pm 0.38$ & $0.058 \pm 0.002$ & X & P &  & 2  \\ 
36624 & $32.1 \pm 0.4$ & $6.65 \pm 2.20$ & $0.054 \pm 0.007$ & D & D & 25.1 & 1  \\ 
37519 & $34.3 \pm 0.5$ & $14.52 \pm 1.96$ & $0.073 \pm 0.009$ & D & Z & 29.0 & 1  \\ 
38257 & $17.3 \pm 0.5$ & $4.38 \pm 0.91$ & $0.136 \pm 0.021$ & C & C &  & 3  \\ 
47969 & $23.5 \pm 1.0$ & $18.73 \pm 4.62$ & $0.061 \pm 0.010$ & D & Z & 16.6 & 1  \\ 
47967 & $20.8 \pm 0.3$ & $9.21 \pm 0.28$ & $0.086 \pm 0.011$ & D & D &  & 2  \\ 
48249 & $19.3 \pm 0.6$ & $10.88 \pm 0.28$ & $0.076 \pm 0.006$ & D & D &  & 2  \\ 
48252 & $ $ & $9.62 \pm 0.32$ & $ $ & D & D &  & 2  \\ 
48373 & $20.3 \pm 0.6$ & $13.28 \pm 4.83$ & $0.090 \pm 0.015$ & D & D & 15.5 & 1  \\ 
48438 & $36.1 \pm 0.3$ & $7.21 \pm 1.78$ & $0.101 \pm 0.014$ & D & D & 29.9 & 1  \\ 
48604 & $14.5 \pm 0.8$ & $9.53 \pm 0.12$ & $0.077 \pm 0.016$ & D & D &  & 4  \\ 
48764 & $28.0 \pm 3.0$ & $11.74 \pm 2.43$ & $0.062 \pm 0.015$ & D & Z & 30.3 & 1  \\ 
51345 & $23.7 \pm 0.7$ & $13.66 \pm 2.16$ & $0.070 \pm 0.007$ & D & Z & 26.4 & 1  \\ 
51346 & $20.8 \pm 1.0$ & $9.46 \pm 1.11$ & $0.078 \pm 0.016$ & D & D &  & 3  \\ 
51354 & $28.2 \pm 0.3$ & $15.85 \pm 3.63$ & $0.080 \pm 0.012$ & D & Z & 21.3 & 1  \\ 
51359 & $22.6 \pm 0.3$ & $12.63 \pm 1.17$ & $0.057 \pm 0.005$ & D & Z &  & 2  \\ 
51935 & $23.3 \pm 1.1$ & $11.54 \pm 1.22$ & $0.068 \pm 0.007$ & D & Z &  & 3  \\ 
51962 & $27.5 \pm 0.7$ & $10.07 \pm 0.20$ & $0.077 \pm 0.014$ & D & D &  & 4  \\ 
51364 & $26.5 \pm 0.7$ & $5.68 \pm 1.81$ & $0.068 \pm 0.009$ & D & D & 29.8 & 1  \\ 
51365 & $42.0 \pm 0.3$ & $9.30 \pm 1.25$ & $0.065 \pm 0.010$ & D & Z & 45.0 & 1  \\ 
51910 & $26.7 \pm 0.3$ & $10.41 \pm 3.75$ & $0.067 \pm 0.008$ & D & D & 15.2 & 1  \\ 
51958 & $27.4 \pm 0.4$ & $11.16 \pm 2.96$ & $0.075 \pm 0.008$ & D & Z & 19.6 & 1  \\ 
51962 & $27.5 \pm 0.7$ & $15.57 \pm 2.78$ & $0.077 \pm 0.014$ & D & Z & 21.6 & 1  \\ 
51994 & $19.8 \pm 2.2$ & $8.56 \pm 2.39$ & $0.060 \pm 0.019$ & D & D & 23.0 & 1  \\ 
52273 & $17.3 \pm 1.2$ & $15.45 \pm 4.77$ & $0.085 \pm 0.019$ & D & Z & 16.0 & 1  \\ 
52767 & $25.1 \pm 0.8$ & $14.49 \pm 5.57$ & $0.045 \pm 0.010$ & D & Z & 13.4 & 1  \\ 
52511 & $21.6 \pm 0.8$ & $12.98 \pm 1.10$ & $0.060 \pm 0.010$ & D & Z &  & 3  \\ 
54596 & $17.5 \pm 0.9$ & $2.17 \pm 1.16$ & $0.057 \pm 0.012$ & C & C &  & 3  \\ 
54656 & $37.7 \pm 0.4$ & $12.78 \pm 0.10$ & $0.073 \pm 0.008$ & D & Z &  & 4  \\ 
55060 & $28.8 \pm 0.4$ & $11.63 \pm 0.16$ & $0.093 \pm 0.011$ & D & Z &  & 4  \\ 
55267 & $24.1 \pm 1.3$ & $11.60 \pm 4.17$ & $0.070 \pm 0.018$ & D & Z & 14.0 & 1  \\ 
55419 & $31.2 \pm 0.3$ & $7.94 \pm 0.06$ & $0.095 \pm 0.007$ & D & D &  & 4  \\ 
55457 & $24.0 \pm 2.8$ & $8.79 \pm 1.02$ & $0.053 \pm 0.017$ & D & D &  & 3  \\ 
55460 & $15.3 \pm 1.1$ & $6.34 \pm 1.48$ & $0.131 \pm 0.032$ & X & X &  & 3  \\ 
55474 & $21.5 \pm 1.0$ & $15.38 \pm 4.46$ & $0.096 \pm 0.022$ & D & Z & 17.0 & 1  \\ 
55678 & $17.6 \pm 1.1$ & $11.01 \pm 1.29$ & $0.052 \pm 0.012$ & D & Z &  & 3  \\ 
55702 & $20.0 \pm 1.1$ & $12.29 \pm 3.76$ & $0.077 \pm 0.016$ & D & Z & 15.0 & 1  \\ 
56968 & $25.7 \pm 0.5$ & $15.86 \pm 0.20$ & $0.066 \pm 0.006$ & D & Z &  & 2  \\ 
56976 & $23.6 \pm 0.4$ & $6.73 \pm 1.63$ & $0.099 \pm 0.013$ & D & D & 32.8 & 1  \\ 
57013 & $24.1 \pm 0.2$ & $10.37 \pm 1.22$ & $0.078 \pm 0.013$ & D & D &  & 3  \\ 
62201 & $19.1 \pm 0.6$ & $8.89 \pm 1.32$ & $0.044 \pm 0.009$ & D & D &  & 3  \\ 
62426 & $25.6 \pm 0.3$ & $7.31 \pm 3.77$ & $0.054 \pm 0.008$ & D & Z & 15.4 & 1  \\ 
63923 & $16.4 \pm 1.5$ & $12.31 \pm 2.89$ & $0.114 \pm 0.033$ & D & D & 20.1 & 1  \\ 
63955 & $22.1 \pm 0.6$ & $12.33 \pm 1.23$ & $0.048 \pm 0.009$ & D & Z &  & 3  \\ 
64270 & $16.5 \pm 0.7$ & $10.60 \pm 1.65$ & $0.049 \pm 0.011$ & D & D &  & 3  \\ 
65590 & $20.3 \pm 1.2$ & $6.91 \pm 1.29$ & $0.052 \pm 0.012$ & X & P &  & 3  \\ 
67548 & $17.9 \pm 2.1$ & $7.17 \pm 3.64$ & $0.096 \pm 0.031$ & D & D & 20.4 & 1  \\ 
68444 & $26.4 \pm 0.5$ & $13.30 \pm 0.07$ & $0.067 \pm 0.004$ & D & Z &  & 4  \\ 
73795 & $20.0 \pm 0.5$ & $6.64 \pm 4.10$ & $0.077 \pm 0.019$ & D & D & 13.4 & 1  \\ 
76804 & $16.6 \pm 0.4$ & $7.29 \pm 0.21$ & $0.122 \pm 0.014$ & D & D &  & 2  \\ 
76820 & $17.5 \pm 0.7$ & $9.29 \pm 1.16$ & $0.070 \pm 0.013$ & D & D &  & 3  \\ 
76837 & $20.0 \pm 1.2$ & $11.31 \pm 1.17$ & $0.070 \pm 0.015$ & D & Z &  & 3  \\ 
76857 & $33.7 \pm 0.4$ & $7.69 \pm 2.26$ & $0.079 \pm 0.011$ & D & D & 24.5 & 1  \\ 
76867 & $44.1 \pm 0.4$ & $11.10 \pm 0.97$ & $0.083 \pm 0.009$ & D & Z & 57.8 & 1  \\ 
77891 & $16.4 \pm 1.4$ & $6.93 \pm 1.54$ & $0.045 \pm 0.014$ & X & P &  & 3  \\ 
82055 & $24.9 \pm 0.6$ & $8.75 \pm 1.53$ & $0.103 \pm 0.020$ & D & Z & 36.2 & 1  \\ 
84709 & $19.5 \pm 1.7$ & $11.64 \pm 0.33$ & $0.088 \pm 0.016$ & D & D &  & 2  \\ 
90380 & $17.9 \pm 1.8$ & $11.98 \pm 3.49$ & $0.055 \pm 0.018$ & D & Z & 16.0 & 1  \\ 
99306 & $16.4 \pm 0.9$ & $11.30 \pm 0.12$ & $0.072 \pm 0.016$ & D & Z &  & 4  \\ 
99328 & $15.5 \pm 0.5$ & $8.28 \pm 0.74$ & $0.116 \pm 0.012$ & D & D &  & 2  \\ 
105685 & $18.9 \pm 0.8$ & $6.54 \pm 0.48$ & $0.049 \pm 0.007$ & D & D &  & 2  \\ 
105746 & $17.0 \pm 0.8$ & $11.08 \pm 1.44$ & $0.088 \pm 0.016$ & D & Z &  & 3  \\ 
105720 & $13.9 \pm 0.8$ & $9.42 \pm 1.07$ & $0.076 \pm 0.016$ & D & D &  & 3  \\ 
105808 & $21.8 \pm 0.4$ & $8.03 \pm 2.28$ & $0.084 \pm 0.013$ & D & D & 24.4 & 1  \\ 
105896 & $13.5 \pm 0.8$ & $5.20 \pm 1.16$ & $0.140 \pm 0.031$ & X & X &  & 3  \\ 
110380 & $15.8 \pm 1.1$ & $13.43 \pm 3.00$ & $0.113 \pm 0.034$ & D & Z & 20.0 & 1  \\ 
111113 & $15.3 \pm 0.7$ & $14.39 \pm 0.31$ & $0.075 \pm 0.011$ & D & D &  & 2  \\ 
114208 & $20.3 \pm 0.6$ & $9.44 \pm 1.24$ & $0.057 \pm 0.014$ & D & D &  & 3  \\ 
120453 & $12.3 \pm 0.3$ & $4.68 \pm 0.69$ & $0.057 \pm 0.004$ & X & P &  & 2  \\ 
120454 & $16.7 \pm 0.5$ & $14.12 \pm 1.43$ & $0.033 \pm 0.006$ & D & Z &  & 3  \\ 
122592 & $ $ & $18.67 \pm 5.35$ & $ $ & D & Z & 14.3 & 1  \\ 
122733 & $15.2 \pm 0.9$ & $14.35 \pm 4.40$ & $0.083 \pm 0.020$ & D & Z & 16.8 & 1  \\ 
124729 & $ $ & $14.78 \pm 0.49$ & $ $ & D & Z &  & 2  \\ 
124985 & $19.2 \pm 0.8$ & $9.61 \pm 1.68$ & $0.052 \pm 0.012$ & D & D &  & 3  \\ 
153500 & $14.5 \pm 0.8$ & $10.81 \pm 1.94$ & $0.077 \pm 0.016$ & D & Z &  & 3  \\ 
182445 & $14.3 \pm 0.8$ & $10.76 \pm 1.49$ & $0.079 \pm 0.014$ & D & Z &  & 3  \\ 
182746 & $21.0 \pm 0.4$ & $9.34 \pm 1.67$ & $0.048 \pm 0.010$ & D & D &  & 3  \\ 
213360 & $13.7 \pm 0.8$ & $10.44 \pm 1.12$ & $0.086 \pm 0.019$ & D & Z &  & 3  \\ 
247351 & $16.0 \pm 0.8$ & $11.76 \pm 1.56$ & $0.083 \pm 0.018$ & D & Z &  & 3  \\ 
    \hline
\end{longtable}
\end{center}


\end{appendix}

\end{document}